\documentclass[aps,prb,twocolumn,showpacs,preprintnumbers,amsmath,amssymb,groupedaddress,superscriptaddress]{revtex4}%

\usepackage{graphicx}
\usepackage{epsfig}
\usepackage{dcolumn}
\usepackage{bm}
\usepackage{longtable}
\usepackage{color} 
\usepackage{multirow}

\begin{document}

\preprint{PREPRINT (\today)}

\title{Superconductivity and magnetism in Rb$_{x}$Fe$_{2-y}$Se$_{2}$: Impact of\\ thermal treatment on mesoscopic phase separation}

\author{S.~Weyeneth}\email{wstephen@physik.uzh.ch}
\affiliation{Physik-Institut der Universit\"at Z\"urich, Winterthurerstrasse 190, CH-8057 Z\"urich, Switzerland}

\author{M.~Bendele\footnote{Current address: Dipartimento di Fisica, Universit\'a di Roma ``La Sapienza''- P.~le Aldo Moro 2, I-00185 Roma, Italy}}
\affiliation{Physik-Institut der Universit\"at Z\"urich, Winterthurerstrasse 190, CH-8057 Z\"urich, Switzerland}
\affiliation{Laboratory for Muon Spin Spectroscopy, Paul Scherrer Institute, CH-5232 Villigen PSI, Switzerland}

\author{F.~von~Rohr}
\affiliation{Physik-Institut der Universit\"at Z\"urich, Winterthurerstrasse 190, CH-8057 Z\"urich, Switzerland}

\author{P.~Dluzewski}
\affiliation{Institute of Physics, Polish Academy of Sciences, Aleja Lotnik\'ow 32/46, PL-02-668 Warsaw, Poland}

\author{R.~Puzniak}
\affiliation{Institute of Physics, Polish Academy of Sciences, Aleja Lotnik\'ow 32/46, PL-02-668 Warsaw, Poland}

\author{A.~Krzton-Maziopa\footnote{On leave from Faculty of Chemistry, Warsaw University of Technology, PL-00-664 Warsaw, Poland}}
\affiliation{Laboratory for Developments and Methods, Paul Scherrer Institute, CH-5232 Villigen PSI, Switzerland}

\author{S.~Bosma}
\affiliation{Physik-Institut der Universit\"at Z\"urich, Winterthurerstrasse 190, CH-8057 Z\"urich, Switzerland}

\author{Z.~Guguchia}
\affiliation{Physik-Institut der Universit\"at Z\"urich, Winterthurerstrasse 190, CH-8057 Z\"urich, Switzerland}

\author{R.~Khasanov}
\affiliation{Laboratory for Muon Spin Spectroscopy, Paul Scherrer Institute, CH-5232 Villigen PSI, Switzerland}

\author{Z.~Shermadini}
\affiliation{Laboratory for Muon Spin Spectroscopy, Paul Scherrer Institute, CH-5232 Villigen PSI, Switzerland}

\author{A.~Amato}
\affiliation{Laboratory for Muon Spin Spectroscopy, Paul Scherrer Institute, CH-5232 Villigen PSI, Switzerland}

\author{E.~Pomjakushina}
\affiliation{Laboratory for Developments and Methods, Paul Scherrer Institute, CH-5232 Villigen PSI, Switzerland}

\author{K.~Conder}
\affiliation{Laboratory for Developments and Methods, Paul Scherrer Institute, CH-5232 Villigen PSI, Switzerland}

\author{A.~Schilling}
\affiliation{Physik-Institut der Universit\"at Z\"urich, Winterthurerstrasse 190, CH-8057 Z\"urich, Switzerland}

\author{H.~Keller}
\affiliation{Physik-Institut der Universit\"at Z\"urich, Winterthurerstrasse 190, CH-8057 Z\"urich, Switzerland}

\begin{abstract}
An extended study of the superconducting and normal-state properties of various as-grown and post-annealed Rb$_{x}$Fe$_{2-y}$Se$_{2}$ single crystals is presented. Magnetization experiments evidence that annealing of Rb$_{x}$Fe$_{2-y}$Se$_{2}$ at 413~K, well below the onset of phase separation $T_{\rm p}\simeq489$~K, neither changes the magnetic nor the superconducting properties of the crystals. In addition, annealing at 563~K, well above $T_{\rm p}$, suppresses the superconducting transition temperature $T_{\rm c}$ and leads to an increase of the antiferromagnetic susceptibility accompanied by the creation of ferromagnetic impurity phases, which are developing with annealing time. However, annealing at $T=488~{\rm K}\simeq T_{\rm p}$ increases $T_{\rm c}$ up to 33.3~K, sharpens the superconducting transition, increases the lower critical field, and strengthens the screening efficiency of the applied magnetic field. Resistivity measurements of the as-grown and optimally annealed samples reveal an increase of the upper critical field along both crystallographic directions as well as its anisotropy. Muon spin rotation and scanning transmission electron microscopy experiments suggest the coexistence of two phases below $T_{\rm p}$: a magnetic majority phase of Rb$_{2}$Fe$_{4}$Se$_{5}$ and a non-magnetic minority phase of Rb$_{0.5}$Fe$_{2}$Se$_{2}$. Both microscopic techniques indicate that annealing the specimens just at $T_{\rm p}$ does not affect the volume fraction of the two phases, although the magnetic field distribution in the samples changes substantially. This suggests that the microstructure of the sample, caused by mesoscopic phase separation, is modified by annealing just at $T_{\rm p}$, leading to an improvement of the superconducting properties of Rb$_{x}$Fe$_{2-y}$Se$_{2}$ and an enhancement of $T_{\rm c}$.
\end{abstract}


\maketitle
\section{Introduction}
Iron-chalcogenide superconductors are usually related to the selenium deficient compound FeSe$_{1-x}$, having a transition temperature $T_{\rm c}\simeq8$~K.\cite{Hsu2009, Pomjakushina2009} Higher $T_{\rm c}$'s can be accessed by applying hydrostatic pressure $p$,\cite{Medvedev2009} by inducing chemical pressure,\cite{Mizuguchi2009, Sales2009} or by intercalating alkali atoms between the Fe$_2$Se$_2$-layers, yielding $A_x$Fe$_{2-y}$Se$_2$ ($A$ = K, Rb, Cs).\cite{Guo2010, Krzton2011, Li2011} Besides superconductivity, many iron-chalcogenides feature coexisting magnetic order, where subtle modifications of the crystal structure lead to drastic changes in superconducting and magnetic properties. This is the case for the compound Rb$_{x}$Fe$_{2-y}$Se$_{2}$, which is superconducting below $T_{\rm c}\simeq33$~K and antiferromagnetic below the N\'eel temperature $T_{\rm N}$ as high as 500~K to 540~K.\cite{Shermadini2011, Liu2011} In addition to these superconducting and magnetic orders, iron-vacancy ordering accompanied by a structural distortion at the temperature $T_{\rm s}$ as well as phase separation in magnetic and nonmagnetic domains at the temperature $T_{\rm p}$ are observed.\cite{Pomjakushin2011} 
\\\indent
Although it was shown by various groups that $A_x$Fe$_{2-y}$Se$_2$ exhibits bulk superconductivity,\cite{Ying2011, Tsurkan2011, Bosma2012} muon spin rotation ($\mu$SR) experiments reported that only a minor volume fraction of $\sim$10\% of the sample is superconducting, whereas $\sim$90\% of the  volume is antiferromagnetic.\cite{Shermadini2012} From neutron experiments the minority phase was identified to have the $I4/mmm$ space group with a small in-plane lattice constant $a$ and a large out-of-plane lattice constant $c$.\cite{Pomjakushin2012} It was discussed whether $A_x$Fe$_{2-y}$Se$_2$ should be treated as a filamentary or granular superconductor.\cite{Shen2011} Besides, mesoscopic phase separation in Rb$_x$Fe$_{2-y}$Se$_2$ was reported to prevail down to the nanoscale.\cite{Ricci2011,Ricci2011_b,  Lei2011_c, Yan2011, Wang2011, Bosak2011} Microscopic techniques probing the stoichiometry of these distinct phases yield in average the composition Rb$_2$Fe$_4$Se$_5$ for the antiferromagnetic vacancy ordered majority phase (245-phase) and the composition Rb$_{1-x}$Fe$_2$Se$_2$ for the superconducting Rb-deficient minority phase (122-phase).\cite{Texier2012, Speller2012} Thus, the studied material may be treated as follows: the minority 122-phase is superconducting and is embedded in an antiferromagnetic matrix of the vacancy ordered 245-phase.
\\\indent
Interestingly, it was observed that some post-annealed iron-chalcogenide samples may become superconducting despite their insulating as-grown behavior.\cite{Ryu2011, Ozaki2011, Lei2011, Han2012} It was discussed that a possible change in the vacancy ordering and the related phase separation might be related to the observed changes in the electronic properties.\cite{Han2012} Obviously, by carefully tuning the conditions of annealing, one may gain direct control of the phase separation in $A_x$Fe$_{2-y}$Se$_2$ and by that of the superconducting and magnetic properties. In order to examine this scenario and to investigate the influence of vacancy ordering and phase separation on superconductivity and magnetism, we performed an extended study of thermally treated Rb$_{x}$Fe$_{2-y}$Se$_2$ single crystals.
\\\indent
\section{Experimental details}
\begin{figure}[b!]
\centering
\includegraphics[width=\linewidth]{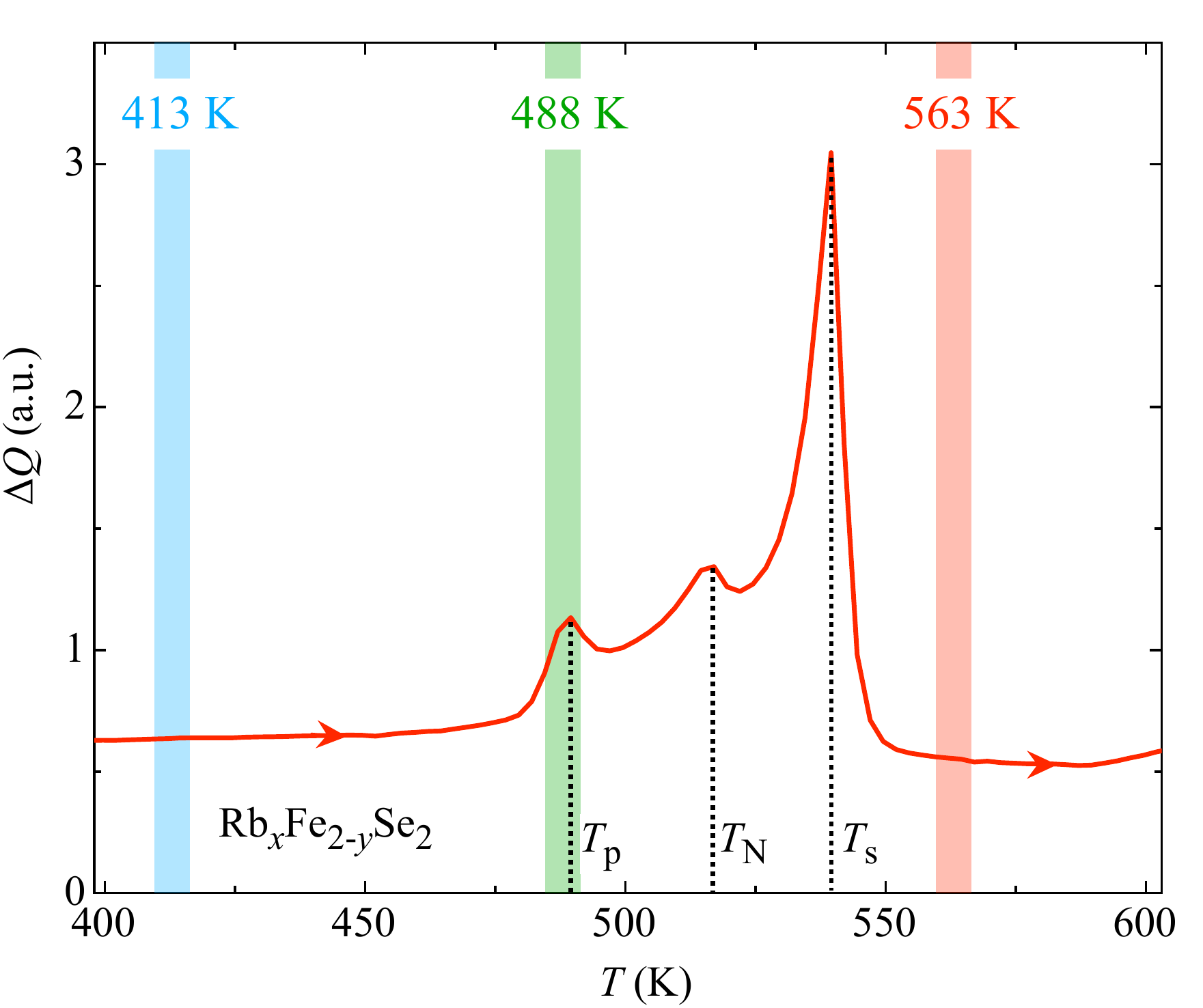}
\caption{(color online) Differential heat $\Delta Q$ for a Rb$_x$Fe$_{2-y}$Se$_2$ single crystal recorded between 400 and 600~K with a constant heating rate of 20 K/min. Three distinct peaks are observed, related to the three onset temperatures $T_{\rm p}\simeq489$~K, $T_{\rm N}\simeq517$~K, and $T_{\rm s}\simeq540$~K (see text). The three annealing temperatures 413~K, 488~K, and 563~K were chosen to post-anneal the as-grown Rb$_x$Fe$_{2-y}$Se$_2$ crystals for the subsequent experiments.}
\label{fig1}
\end{figure}
A set of Rb$_x$Fe$_{2-y}$Se$_2$ single crystals with nominal composition Rb$_{0.85}$Fe$_{1.90}$Se$_2$ was grown by the Bridgman method, similarly as described in Refs.~\onlinecite{Krzton2011} and \onlinecite{Krzton2012}. Here, a mixture of high purity Fe, Se, and Rb (at least 99.99\%; Alfa Aesar) was sealed in an evacuated quartz ampoule. This ampoule, protected by a surrounding evacuated quartz tube, was heated to $1030~^\circ$C for 2~h. The melt was cooled first with $-6~^\circ$C/h to $750~^\circ$C and finally to room temperature at a fast rate of $-200~^\circ$C/h. After synthesis the ampoule was transferred to a glove box and opened there to protect the crystals from degradation in air.
\\\indent 
\begin{figure}[t!]
\centering
\includegraphics[width=\linewidth]{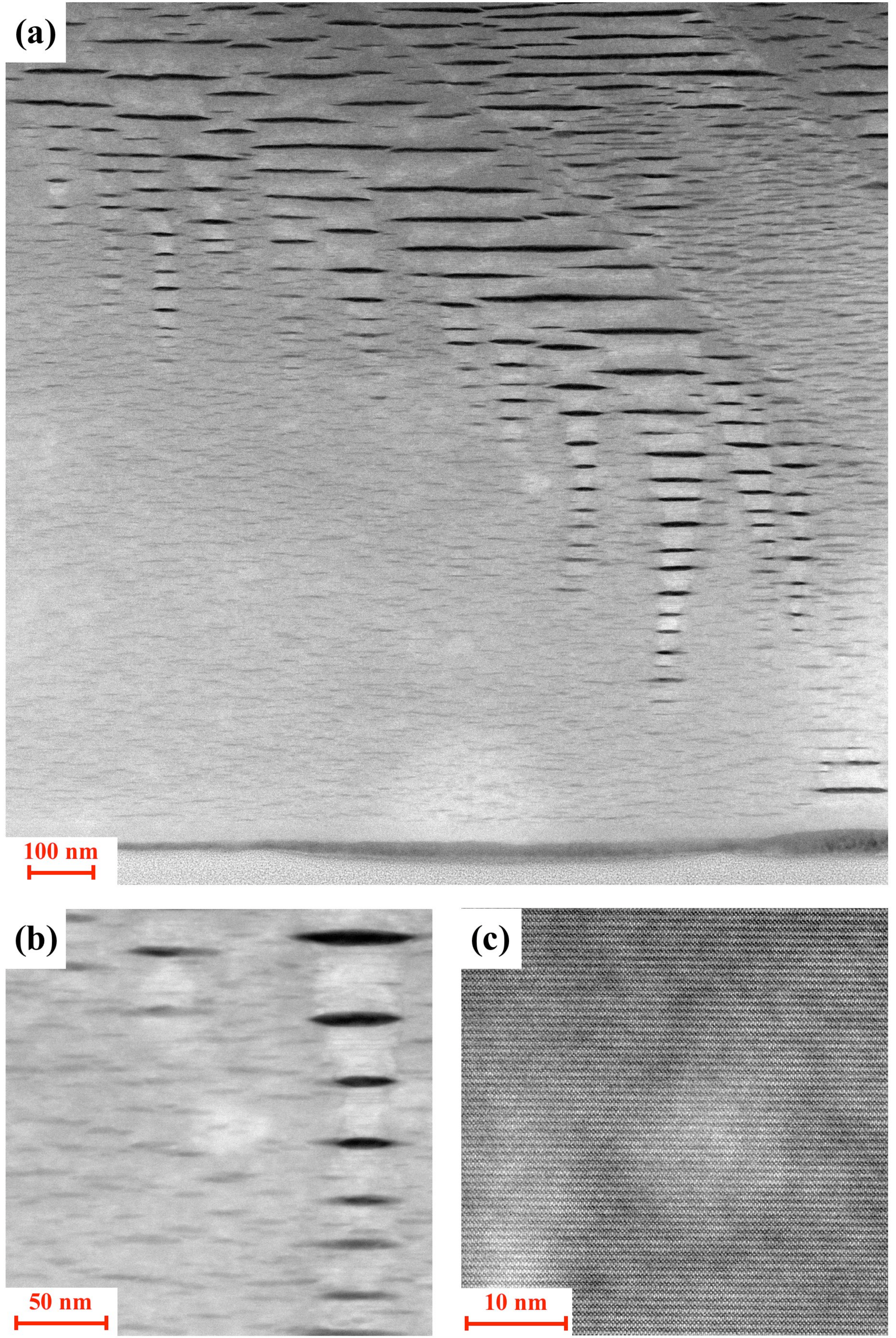}
\caption{(color online) STEM images of as-grown Rb$_x$Fe$_{2-y}$Se$_2$ single crystal taken with the direction of the electron beam perpendicular to the tetragonal $c$-axis. Picture (a) was taken on a square of $\sim1.5\times1.5~\mu$m$^2$, (b) on a square of $\sim250\times250$~nm$^2$, (c) on a square of $\sim50\times50$~nm$^2$. The atomic composition of the darker regions was found to correspond to Rb$_{0.5}$Fe$_{2}$Se$_{2}$, whereas in the brighter regions, the composition is Fe and Rb deficient Rb$_{0.4}$Fe$_{1.6}$Se$_{2}$.}
\label{fig2}
\end{figure}
\begin{table}[t!]
\caption{List of all as-grown and annealed Rb$_x$Fe$_{2-y}$Se$_2$ single-crystal samples investigated by various experimental techniques in this work. The samples with almost identical $T_{\rm c}$ were annealed at a certain temperature $T_{\rm ann}$ during a certain time $t_{\rm ann}$. The as-grown samples are those with $t_{\rm ann}=0$~h. The sample exhibiting the highest $T_{\rm c}$ among the as-grown crystals was annealed at 488~K and is named as A$^*_{488}[t_{\rm ann}]$.}
\label{table0}
\begin{tabular}{p{25mm} c c c }
\hline\hline
Sample						& $T_{\rm ann}$			 		& $t_{\rm ann}$ 		& Experiment		\\\hline\hline
A$_{413}[0~{\rm h}]$			& 413~K							& 0~h 				& magnetometry	\\
A$_{413}[3~{\rm h}]$			& 413~K							& 3~h 				& magnetometry	\\
A$_{413}[36~{\rm h}]$			& 413~K							& 36~h 				& magnetometry	\\\hline
A$_{488}[0~{\rm h}]$			& 488~K							& 0~h 				& magnetometry	\\
A$_{488}[3~{\rm h}]$			& 488~K							& 3~h 			  	& magnetometry	\\
A$_{488}[36~{\rm h}]$			& 488~K							& 36~h 				& magnetometry  	\\\hline
A$_{563}[0~{\rm h}]$			& 563~K							& 0~h 				& magnetometry	\\
A$_{563}[3~{\rm h}]$			& 563~K							& 3~h 				& magnetometry  	\\
A$_{563}[36~{\rm h}]$			& 563~K							& 36~h 				& magnetometry  	\\\hline
A$^*_{488}[0~{\rm h}]$			& 488~K							& 0~h 				& magnetometry	\\
A$^*_{488}[3~{\rm h}]$			& 488~K							& 3~h 				& magnetometry  	\\
A$^*_{488}[36~{\rm h}]$			& 488~K							& 36~h 				& magnetometry  	\\\hline
B$_{488}[0~{\rm h}]$ 			& 488~K							& 0~h 				& transport		\\
B$_{488}[3~{\rm h}]$			& 488~K							& 3~h 				& transport		\\\hline
C$_{488}[0~{\rm h}]$ 			& 488~K							& 0~h 				& $\mu$SR		\\
C$_{488}[60~{\rm h}]$			& 488~K							& 60~h 				& $\mu$SR		\\\hline
D$_{488}[0~{\rm h}]$ 			& 488~K							& 0~h 				& STEM			\\
D$_{488}[3~{\rm h}]$			& 488~K							& 3~h 				& STEM			\\\hline\hline
\end{tabular}
\end{table}
In order to study the thermal evolution of the mesoscopic phase separation, an as-grown Rb$_x$Fe$_{2-y}$Se$_2$ single crystal was initially characterized by differential scanning calorimetry (DSC). With DSC, the differential amount of heat $\Delta Q$ required to increase the sample temperature $T$ by $\Delta T$ with respect to a reference is recorded.\cite{Wunderlich1990} Measurements were performed with a \emph{Netzsch} DSC 204F1 system, by heating up from 290~K to 670~K with a constant heating rate of 20~K/min. Both, sample and reference were always maintained at the same temperature throughout the experiment. In Fig.~\ref{fig1} the measured $\Delta Q$ in the temperature range between 400~K and 600~K for the as-grown single crystal is presented. The three peaks at the temperatures $T_{\rm s}$, $T_{\rm N}$, and $T_{\rm p}$ are related to three distinct onset temperatures of this system: (i) $T_{\rm s}\simeq540$~K corresponds to the onset temperature of iron vacancy ordering, at which the unit cell transforms from the high-temperature $I4/mmm$ structure into a low-temperature superstructure $I4/m$, (ii) $T_{\rm N}\simeq517$~K is the N\'eel temperature, and (iii) $T_{\rm p}\simeq489$~K corresponds to the onset temperature of phase separation between coexisting $I4/mmm$ and $I4/m$ phases.\cite{Pomjakushin2012} 
\\\indent
The mesoscopic phase separation of as-grown Rb$_x$Fe$_{2-y}$Se$_2$ is visualized with scanning transmission electron microscopy (STEM) at room temperature using a {\it Titan} 80-300 {\it Cubed} instrument operating at $300$~keV. The specimens for STEM investigations were carefully prepared by a focused ion beam (FIB) to avoid degradation on air exposition. The STEM images taken with the electron beam perpendicular to the tetragonal $c$-axis are shown in Fig.~\ref{fig2}. The brightness of the STEM images allows to distinguish the actual composition of the sample. According to the results of energy dispersive x-ray spectroscopy (EDXS) the composition of the darker and brighter regions is Rb$_{0.5}$Fe$_{2}$Se$_{2}$ and Rb$_{0.4}$Fe$_{1.6}$Se$_{2}$, respectively. 
\\\indent
Although the transition temperatures $T_{\rm N}$ and $T_{\rm s}$ both correspond to thermodynamic ordering phenomena in this system, the onset of phase separation $T_{\rm p}$ is of different origin. It can be presumed that thermal history of this material crucially influences the phase separation in the sample. This rises the question whether it might be possible to tune the phase separation in Rb$_x$Fe$_{2-y}$Se$_2$ by proper thermal treatment, and by that to control the superconducting and magnetic properties. In order to study the influence of post annealing on the properties of Rb$_x$Fe$_{2-y}$Se$_2$ single crystals, a set of samples was annealed with an \emph{Elite Thermal Systems Ltd.} single zone high temperature furnace at three annealing temperatures characteristic for the studied samples (see Fig.~\ref{fig1}): (i) $T\simeq413$~K (well below $T_{\rm p}$), (ii) $T\simeq488$~K (just at $T_{\rm p}$), and (iii) $T\simeq563$~K (well above $T_{\rm p}$). For this purpose the samples were loaded in a furnace, which was heated from room temperature with a fast rate of $\sim10$~K/min. Having reached the desired annealing temperature $T_{\rm ann}$, the temperature was kept constant for a time $t_{\rm ann}$, after which the samples were removed from the hot furnace and were rapidly cooled back to room temperature. As-grown and annealed samples were systematically studied by various experimental methods. The superconducting and normal-state magnetization was studied with a \emph{Quantum Design} Magnetic Property Measurement System (MPMS) XL with a differential superconducting quantum interference device (SQUID) equipped with a reciprocating sample option (RSO). In order to prevent these samples from degradation in air, all investigated crystals were vacuum sealed in quartz ampoules of $5$~mm diameter and approximately $10$~cm length. The plate-like crystals were oriented with their crystallographic $c$-axis along the ampoule axis and were fixed between two quartz cylinders of approximately $5$~cm length. The diameter of the crystals was adapted to the inner diameter of the quartz tube. Such sample mounting provides a homogenous surrounding of the examined crystal and produces only a minor background signal during the measurements. Resistivity measurements with electrical current flowing in the $ab$-plane were performed with a \emph{Quantum Design} Physical Property Measurement System (PPMS). The Rb$_x$Fe$_{2-y}$Se$_2$ single crystal was cleaved along the $ab$-plane in argon atmosphere inside a glove box and contacted on the cleaved surface by the four-probe technique with gold wires (50~$\mu$m diameter) and silver epoxy. The as-grown sample was sealed directly after the initial measurements inside a quartz ampoule and was subsequently annealed and remeasured. By this procedure we ensured the measurement geometry to stay exactly the same for all the measurements. The $\mu$SR investigations with magnetic fields applied along the $c$-axis were performed with the General Purpose Surface (GPS) $\mu$SR Instrument located at the $\pi$M3 beam line at the Swiss Muon Source (S$\mu$S) at the Paul Scherrer Institute. The $\mu$SR time spectra have been analyzed using the free software package MUSRFIT.\cite{Suter} STEM measurements were done as described above. A list of the various as-grown and annealed samples studied in this work is presented in Table~\ref{table0}.
\\\indent
\section{Results}
\begin{figure}[t!]
\centering
\includegraphics[width=\linewidth]{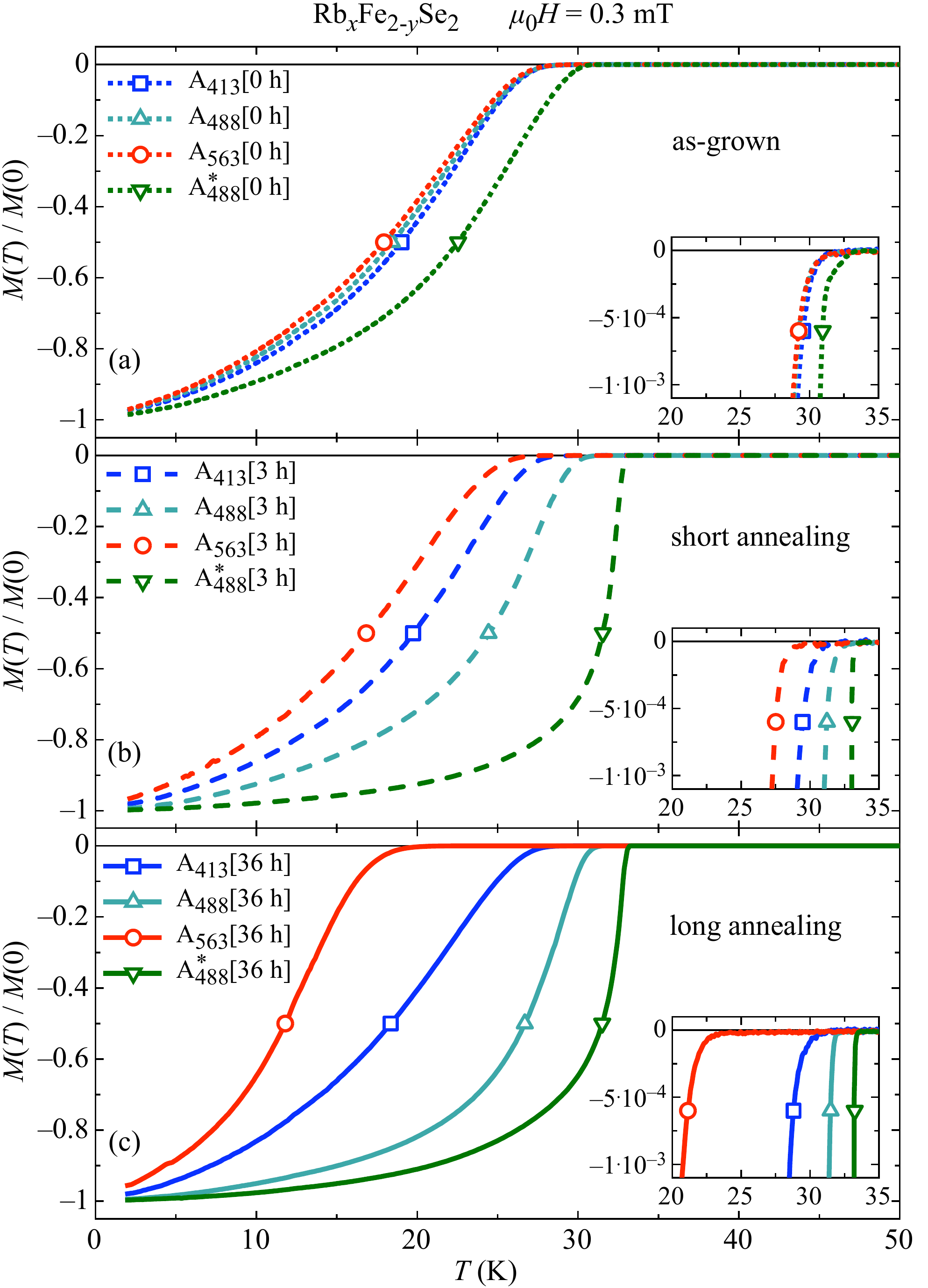}
\caption{(color online) Normalized zfc magnetization $M(T)/M(0)$ for the Rb$_x$Fe$_{2-y}$Se$_2$ single crystals A$_{413}[t_{\rm ann}]$, A$_{488}[t_{\rm ann}]$, A$_{563}[t_{\rm ann}]$, and A$^*_{488}[t_{\rm ann}]$ in a magnetic field $\mu_0 H=0.3$~mT applied along the $c$-axis. The panels present the data for the as-grown samples with $t_{\rm ann}=0$~h (a), annealed samples for $t_{\rm ann}=3$~h (b), and for $t_{\rm ann}=36$~h (c). The respective insets show close-ups of the onset of diamagnetism.}
\label{fig3}
\end{figure}
\begin{figure}[t!]
\centering
\includegraphics[width=\linewidth]{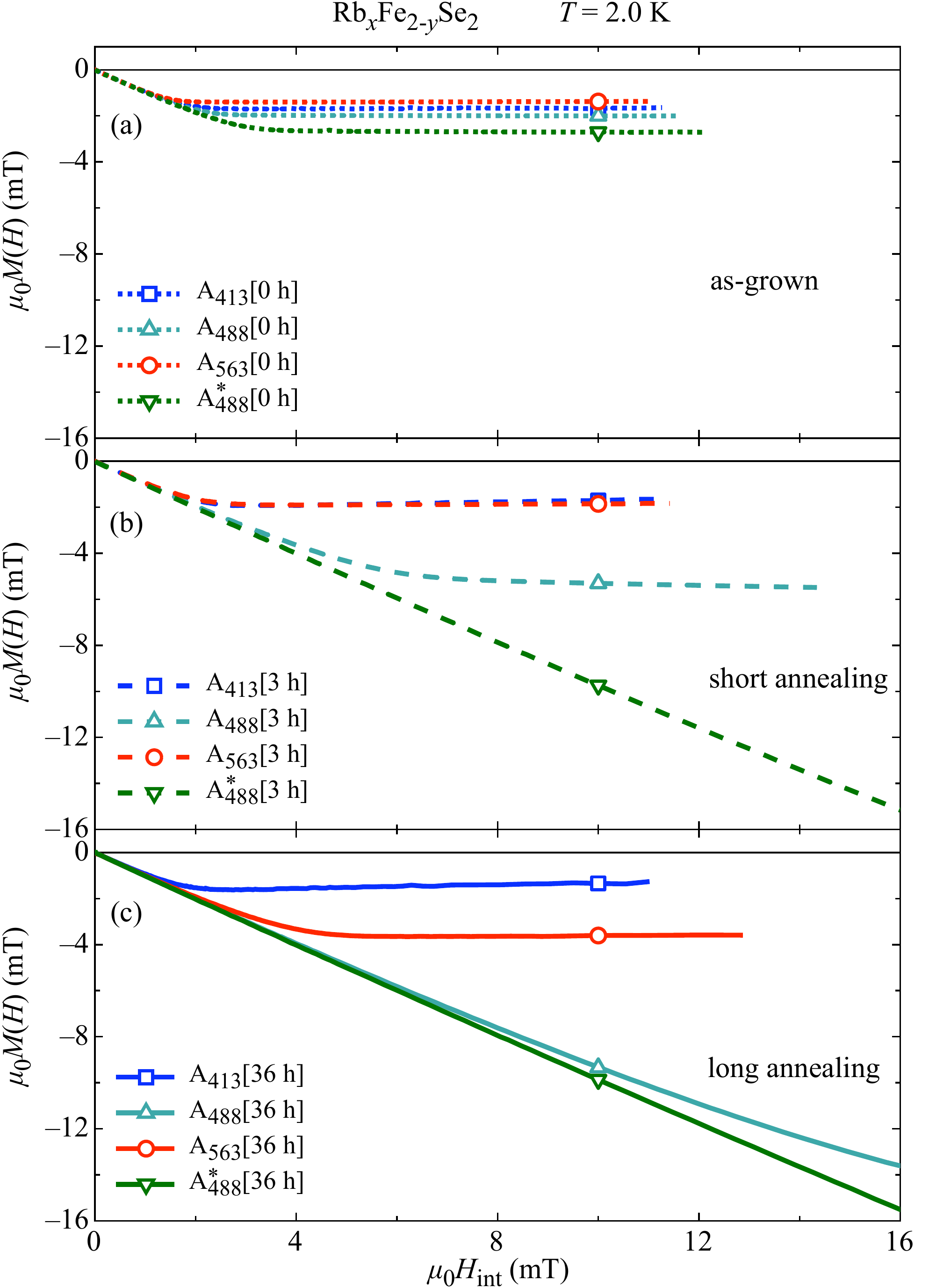}
\caption{(color online) Zero-field cooled (zfc) magnetization curves for the Rb$_x$Fe$_{2-y}$Se$_2$ single crystals A$_{413}[t_{\rm ann}]$, A$_{488}[t_{\rm ann}]$, A$_{563}[t_{\rm ann}]$, and A$^*_{488}[t_{\rm ann}]$ measured at 2.0~K as a function of $H_{\rm int}$ along the $c$-axis. The corresponding $t_{\rm ann}$ of the different panels are $t_{\rm ann}=0$~h (a), $t_{\rm ann}=3$~h (b), and $t_{\rm ann}=36$~h (c). }
\label{fig4}
\end{figure}
In Fig.~\ref{fig3} the zero-field cooled (zfc) magnetization, measured in a magnetic field $\mu_0H=0.3~{\rm mT}$ applied along the $c$-axis for the samples A$_{413}[t_{\rm ann}]$, A$_{488}[t_{\rm ann}]$, A$_{563}[t_{\rm ann}]$, and A$^*_{488}[t_{\rm ann}]$ (see Table~\ref{table0}) with $t_{\rm ann}=0$, 3, and 36~h are shown. The magnetization $M$ was normalized to the individual linearly extrapolated value of $M(0)$. This allows us to directly compare the curves of the various crystals to each other despite their different masses and shapes. In a first step the properties of the pristine as-grown samples ({\it i.e.}, for $t_{\rm ann}=0$~h) were investigated [see Fig.~\ref{fig3}(a)]. After these measurements, the samples were annealed at $T_{\rm ann}$ for 3~h and were remeasured afterwards [see Fig.~\ref{fig3}(b)], then again annealed at $T_{\rm ann}$ for another 33~h (leading to a total annealing time of $t_{\rm ann}=36$~h), and finally remeasured [see Fig.~\ref{fig3}(c)]. During all the measurements and annealings the samples were kept inside the sealed ampoules. The as-grown samples A$_{413}[0~{\rm h}]$, A$_{488}[0~{\rm h}]$, and A$_{563}[0~{\rm h}]$ show very similar behavior, exhibiting superconducting diamagnetism with a rather broad transition width. Only the sample A$^*_{488}[0~{\rm h}]$ exhibits a slightly higher $T_{\rm c}$ and a narrower transition width. The insets to Fig.~\ref{fig3} present a close-up of the onset of diamagnetism. Importantly, the transition temperature $T_{\rm c}$ clearly changes for most of the samples after annealing for $t_{\rm ann}=3$~h and for $t_{\rm ann}=36$~h. Only $T_{\rm c}$ for the sample A$_{413}[t_{\rm ann}]$ is essentially independent of $t_{\rm ann}$. Note that both samples A$_{488}[36~{\rm h}]$ and A$^*_{488}[36~{\rm h}]$ exhibit a clearly narrower transition to the superconducting state with a higher $T_{\rm c}$. In contrast, sample A$_{563}[36~{\rm h}]$ behaves in the opposite way, showing a drastically lower $T_{\rm c}$. The transition width $\Delta T_{\rm c}$ was defined as the inverse of the maximal slope of the normalized magnetization $M/M(0)$ as a function of $T$: 
\begin{equation}
\label{eq1}
\Delta T_{\rm c}=\left(\frac{1}{M(0)}\cdot{\rm max}\left[\frac{dM}{dT}\right]\right)^{-1}.
\end{equation}
The estimated values for $T_{\rm c}$ and $\Delta T_{\rm c}$ for all the samples studied are listed in Table~\ref{table1}. In order to better specify the change for a measured property $P$ with annealing time $t_{\rm ann}$, we introduce the following quantity:
\begin{equation}
\label{eq2}
\delta_{t_{\rm ann}}(P)=\frac{P(t_{\rm ann})-P(0~{\rm h})}{P(0~{\rm h})}.
\end{equation}
With this formula, a clear increase of $T_{\rm c}$ by $\sim5-6\%$ is found for the samples A$_{488}[36~{\rm h}]$ and A$^*_{488}[36~{\rm h}]$ in comparison to the as-grown specimens (see Table~\ref{table1}), whereas $T_{\rm c}$ decreases for sample A$_{563}[36~{\rm h}]$ by $\simeq27.3\%$ and remains almost constant for sample A$_{413}[36~{\rm h}]$. The relative transition width $\Delta T_{\rm c}/T_{\rm c}$ of the samples A$_{413}[t_{\rm ann}]$ and A$_{563}[t_{\rm ann}]$ changes only slightly with annealing, whereas for the samples A$_{488}[t_{\rm ann}]$ and A$^*_{488}[t_{\rm ann}]$ a clear improvement is seen. Note that the transition for A$^*_{488}[t_{\rm ann}]$ becomes almost ideally sharp with long annealing.
\\\indent
\begin{table}[t!]
\caption{Evolution of the transition temperature $T_{\rm c}$ and transition width $\Delta T_{\rm c}$ [see Eq.~(\ref{eq1})] of the samples A$_{413}[t_{\rm ann}]$, A$_{488}[t_{\rm ann}]$, A$_{563}[t_{\rm ann}]$, and A$^*_{488}[t_{\rm ann}]$ with annealing time $t_{\rm ann}$. The changes with annealing $\delta_{t_{\rm ann}}(T_{\rm c})$ and $\delta_{t_{\rm ann}}(\Delta T_{\rm c})$ were calculated applying Eq.~(\ref{eq2}).}
\label{table1}
\begin{tabular}{p{20mm} c c c c c}
\hline\hline
Sample						& $T_{\rm c}$		& $\delta_{t_{\rm ann}}(T_{\rm c})$ 	& $\Delta T_{\rm c}$	& $\delta_{t_{\rm ann}}(\Delta T_{\rm c})$		& $\Delta T_{\rm c}/T_{\rm c}$	\\
							& (K)				& 				& (K)				& 						&  						\\\hline\hline
A$_{413}[0~{\rm h}]$			& 30.1(1)			&				& 14(1)			&						& 47(2)\%					\\
A$_{413}[3~{\rm h}]$			& 30.1(1)			& $\pm0.0$\%		& 14(1)			& $\pm0$\%				& 47(2)\%					\\
A$_{413}[36~{\rm h}]$			& 29.5(1)			& $-$2.0\%		& 16(1)			& +14\%					& 54(2)\%					\\\hline
A$_{488}[0~{\rm h}]$			& 30.0(1)			&				& 16(1) 			& 						& 53(3)\%					\\
A$_{488}[3~{\rm h}]$			& 31.7(1)			& +5.7\%			& 9.5(5)			& $-$41\%				& 30(1)\%					\\
A$_{488}[36~{\rm h}]$			& 31.8(1)			& +6.0\%			& 7.0(4)			& $-$56\%				& 22(1)\%					\\\hline
A$_{563}[0~{\rm h}]$			& 30.0(1)			&				& 17(1)			&						& 57(3)\%					\\
A$_{563}[3~{\rm h}]$			& 28.0(1)			& $-$6.7\%		& 15(1)			& $-$12\%				& 54(3)\%					\\
A$_{563}[36~{\rm h}]$ 			& 21.8(1)			& $-$27.3\%		& 10(1)			& $-$41\%				& 46(4)\%					\\\hline
A$^*_{488}[0~{\rm h}]$			& 31.6(1)			&				& 13(1)			&						& 41(2)\%					\\
A$^*_{488}[3~{\rm h}]$			& 33.1(1)			& +4.7\%			& 2.2(2)			& $-$83\%				& 6.6(3)\%				\\
A$^*_{488}[36~{\rm h}]$			& 33.3(1)			& +5.4\%			& 2.1(2)			& $-$84\%				& 6.3(3)\%				\\
\hline\hline								
\end{tabular}
\end{table}
\begin{table}[t!]
\caption{Evolution of the superconducting susceptibility $\chi_{\rm sc}(\mu_0H_{\rm int})$ [see Eq.~(\ref{eq5})] of the samples A$_{413}[t_{\rm ann}]$, A$_{488}[t_{\rm ann}]$, A$_{563}[t_{\rm ann}]$, and A$^*_{488}[t_{\rm ann}]$ with annealing time $t_{\rm ann}$.}
\label{table2}
\begin{tabular}{p{25mm} c c c c c c}
\hline\hline
Sample							& $\chi_{\rm sc}(1~{\rm mT})$	& $\chi_{\rm sc}(10~{\rm mT})$	\\
								& 						&  						\\\hline\hline
A$_{413}[0~{\rm h}]$				& -0.954(6)				& -0.172(8)				\\
A$_{413}[3~{\rm h}]$				& -0.962(5)				& -0.174(9)				\\
A$_{413}[36~{\rm h}]$				& -0.920(7)				& -0.134(5)				\\\hline
A$_{488}[0~{\rm h}]$				& -0.915(7)				& -0.175(9)				\\
A$_{488}[3~{\rm h}]$				& -0.955(9)				& -0.453(9)				\\
A$_{488}[36~{\rm h}]$				& -0.984(3)				& -0.881(4)				\\\hline
A$_{563}[0~{\rm h}]$				& -0.906(3)				& -0.124(6)				\\
A$_{563}[3~{\rm h}]$				& -0.908(6)				& -0.162(8)				\\
A$_{563}[36~{\rm h}]$				& -0.977(4)				& -0.322(9)				\\\hline
A$^*_{488}[0~{\rm h}]$				& -0.976(3)				& -0.240(9)				\\
A$^*_{488}[3~{\rm h}]$				& -0.989(2)				& -0.940(2)				\\
A$^*_{488}[36~{\rm h}]$				& -0.990(2)				& -0.954(2)				\\
\hline\hline																					
\end{tabular}
\end{table}
\begin{figure}[t!]
\centering
\includegraphics[width=\linewidth]{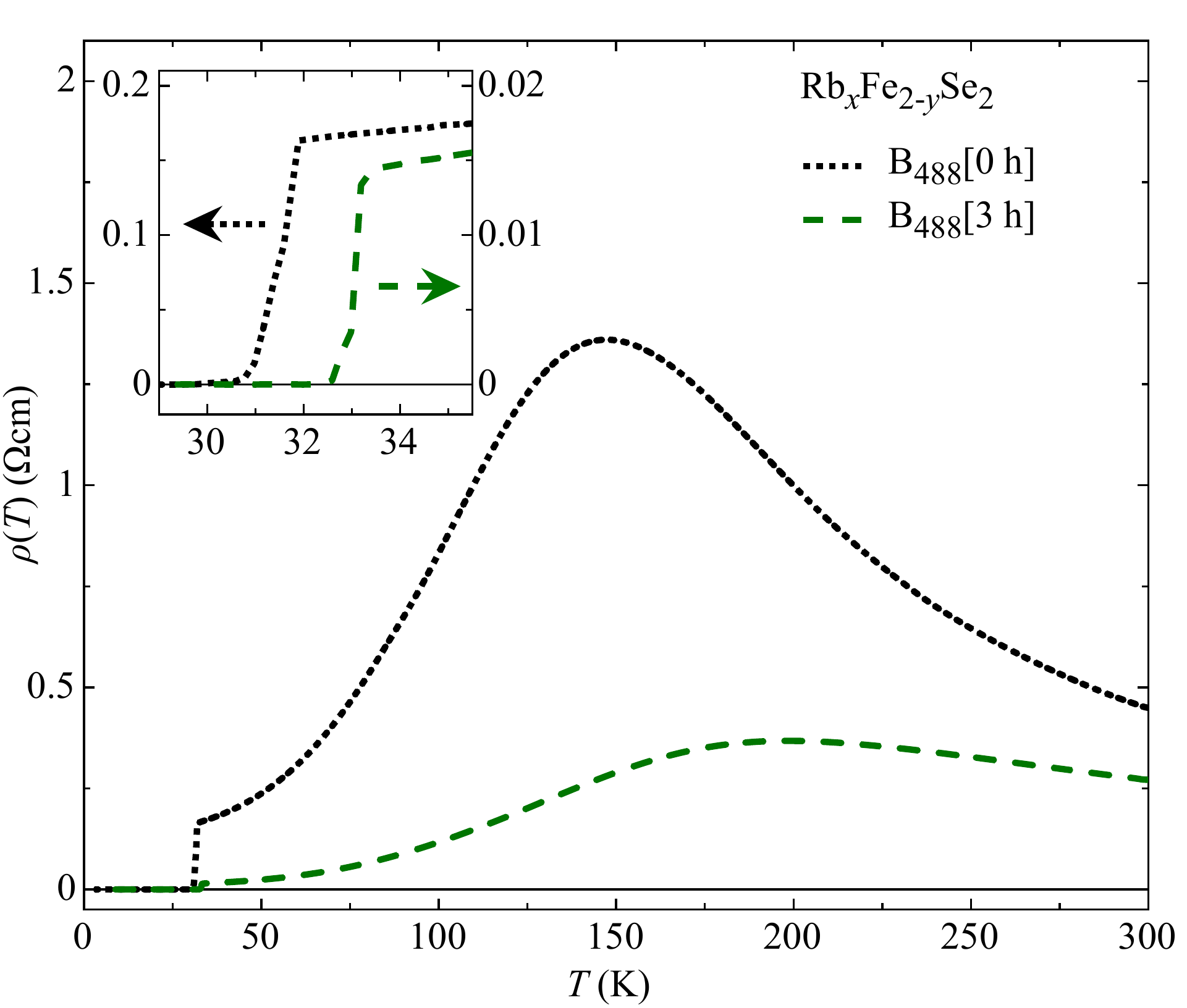}
\caption{(color online) In-plane resistivity $\rho$ of the Rb$_x$Fe$_{2-y}$Se$_2$ samples B$_{488}[0~{\rm h}]$ and B$_{488}[3~{\rm h}]$. The pronounced hump in the normal-state resistivity of the as-grown sample B$_{488}[0~{\rm h}]$ decreases dramatically after annealing and the superconducting $T_{\rm c}$ increases from 31.5~K to 33.1~K.} 
\label{fig5}
\end{figure}
\begin{figure}[t!]
\centering
\includegraphics[width=\linewidth]{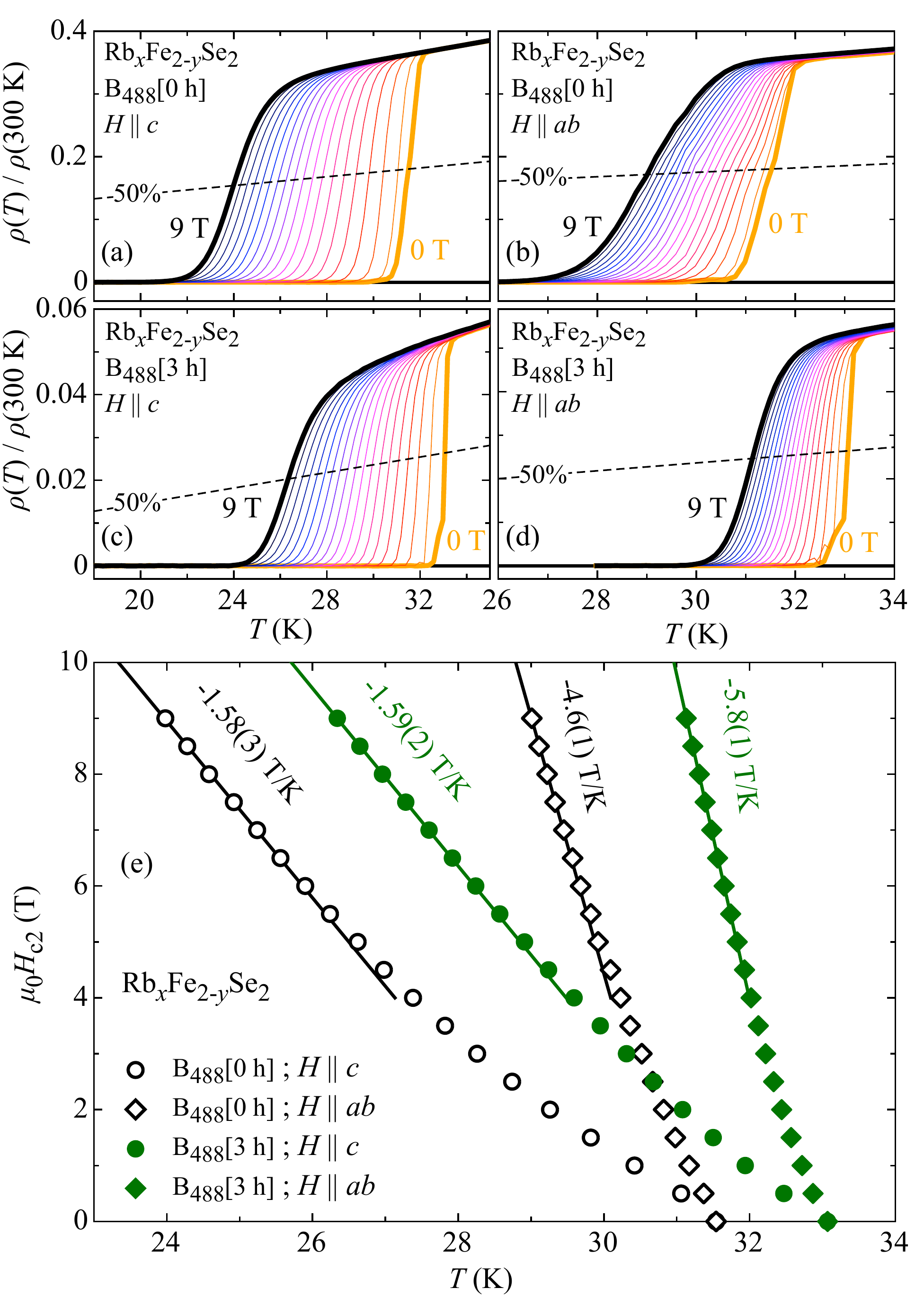}
\caption{(color online) Resistivity of the samples B$_{488}[t_{\rm ann}]$ for magnetic fields between 0 and 9~T, varied by 0.5~T steps, for fields in the $ab$-plane and along the $c$-axis. The measurements were performed for the as-grown sample B$_{488}[0~{\rm h}]$ [panels (a) and (b)] and for the annealed sample B$_{488}[3~{\rm h}]$ [panels (c) and (d)], with $H$ applied along the $c$-axis and in the $ab$-plane. The dashed lines denotes 50~\% of the extrapolated normal-state resistivity, which was used as a criterion to determine $H_{\rm c2}(T)$, shown in panel (e). The transition temperature $T_{\rm c}$ increases by 1.6~K as a result of annealing. The solid lines are guides to the linear part of the $H_{\rm c2}(T)$ curves, used in the WHH-approximation [Eq.~(\ref{eq6})].} 
\label{fig6}
\end{figure}
\begin{figure}[t!]
\centering
\includegraphics[width=\linewidth]{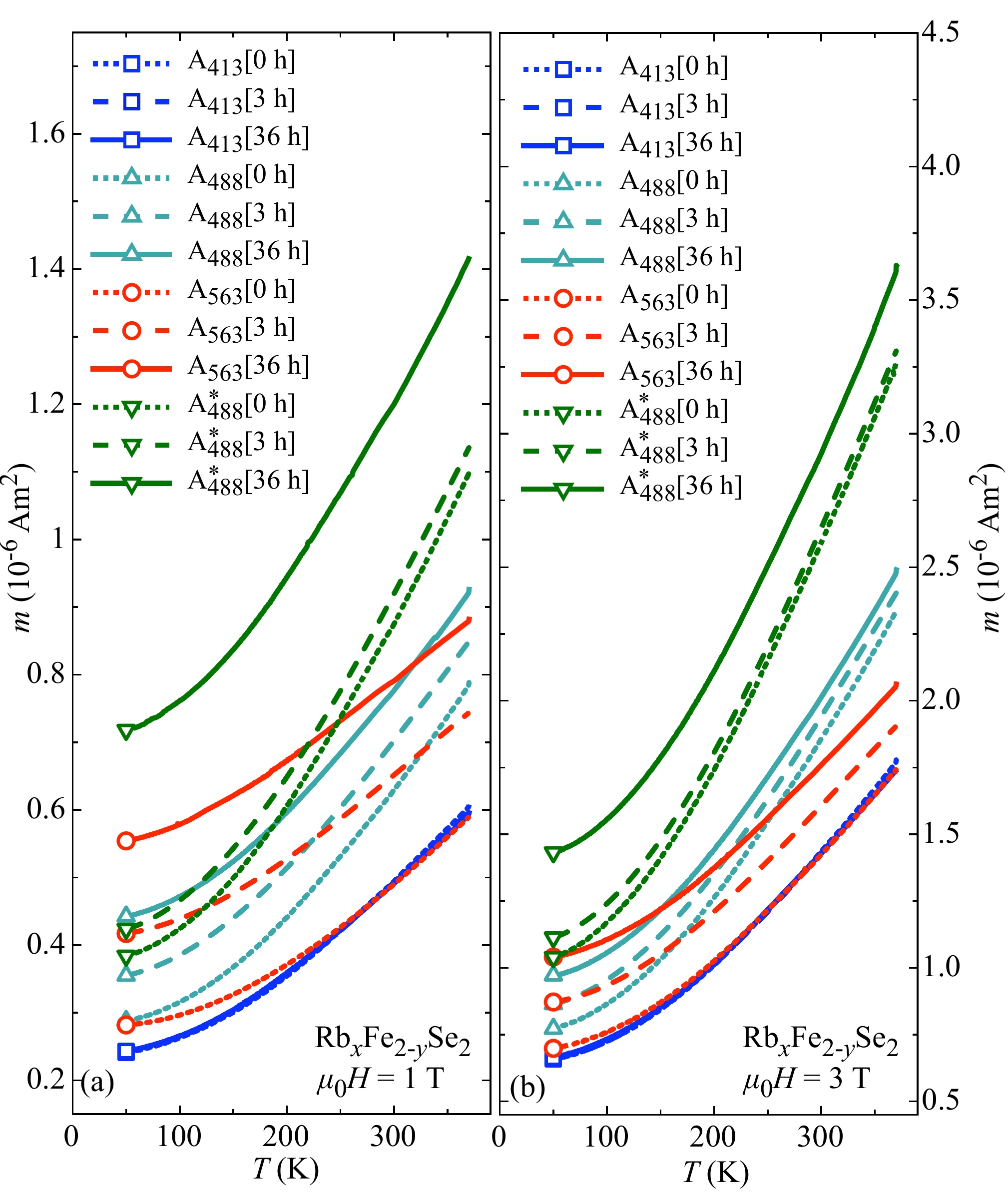}
\caption{(color online) Measured magnetic moment $m(T)$ of pristine and annealed Rb$_x$Fe$_{2-y}$Se$_2$ single crystals in the temperature range between 50 and 370~K for magnetic fields of 1~T (a) and 3~T (b), applied along the $c$-axis.}
\label{fig7}
\end{figure}
\begin{figure}[t!]
\centering
\includegraphics[width=\linewidth]{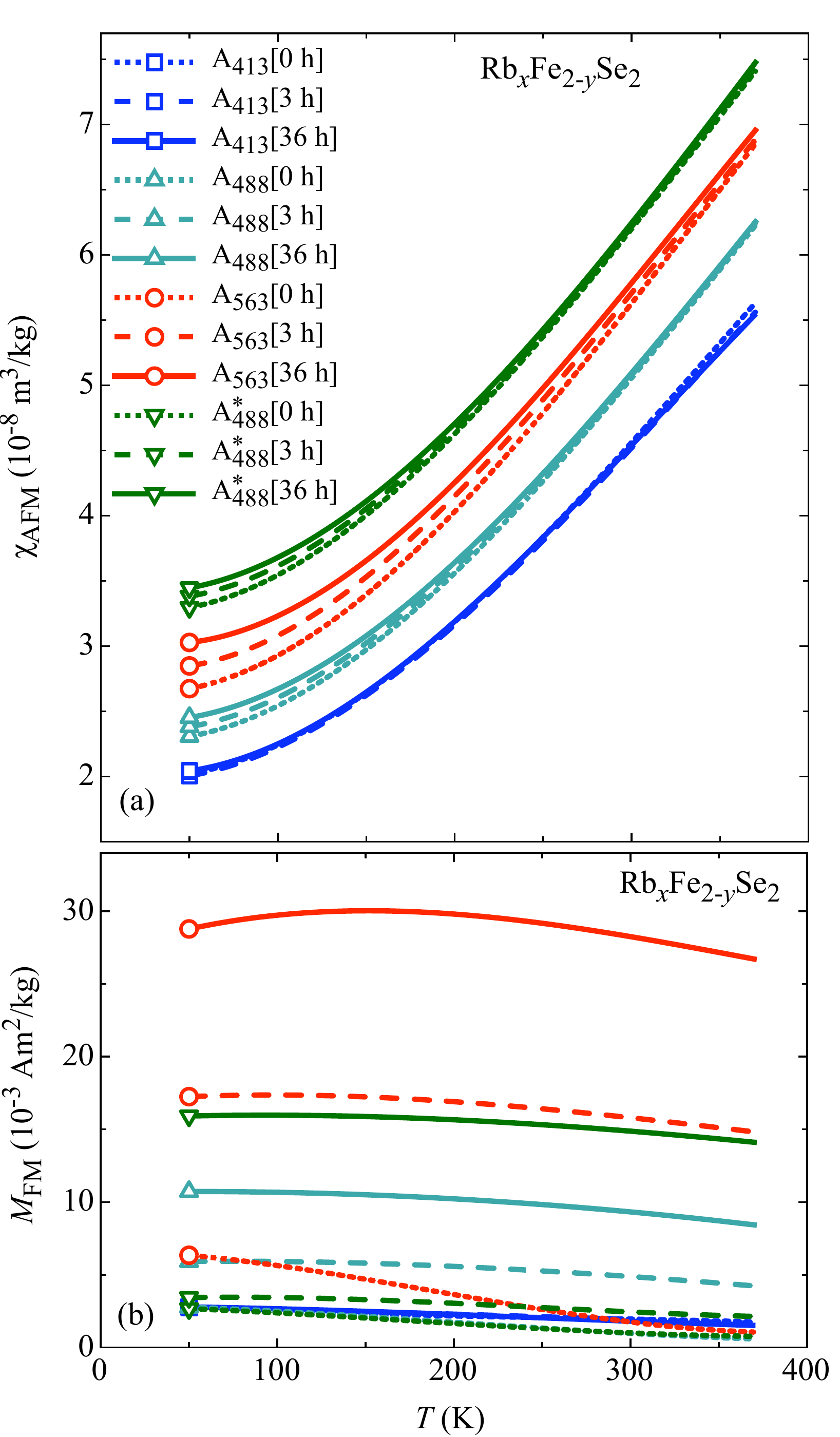}
\caption{(color online) (a) Antiferromagnetic susceptibility $\chi_{\rm AFM}(T)$ in the normal state of pristine and annealed Rb$_x$Fe$_{2-y}$Se$_2$ single crystals determined from the data shown in Fig.~\ref{fig7} using Eq.~(\ref{eq8}). For clarity, the curves representing the four different annealing sets are vertically shifted to each other. Whereas no change of $\chi_{\rm AFM}(T)$ is found by annealing for sample A$_{413}[t_{\rm ann}]$, for all other samples $\chi_{\rm AFM}(T)$ increases with increasing $t_{\rm ann}$. (b) Ferromagnetic component $M_{\rm FM}(T)$, being constant for sample A$_{413}[t_{\rm ann}]$ as a function of $t_{\rm ann}$. For all other samples $M_{\rm FM}(50~{\rm K})$ increases substantially with increasing $t_{\rm ann}$. }
\label{fig8}
\end{figure}
\begin{table*}[t!]
\caption{Evolution of $T_{\rm c}$, $-dH_{\rm c2}^{||c}/dT$, and $-dH_{\rm c2}^{||ab}/dT$ with annealing time $t_{\rm ann}$ for fields applied parallel to the $c$-axis and to the $ab$-plane for samples B$_{488}[0~{\rm h}]$ and B$_{488}[3~{\rm h}]$. The changes with annealing $\delta_{t_{\rm ann}}(T_{\rm c})$, $\delta_{t_{\rm ann}}(dH_{\rm c2}^{||c}/dT)$, and $\delta_{t_{\rm ann}}(dH_{\rm c2}^{||ab}/dT)$ were calculated applying Eq.~(\ref{eq2}).}
\label{table3}
\begin{tabular}{ p{30mm} c c c c c c c c c c c c}
\hline\hline	
Sample						& $T_{\rm c}$ 	& $\delta_{t_{\rm ann}}(T_{\rm c})$		& $-\mu_0dH_{\rm c2}^{||c}/dT$	& $\delta_{t_{\rm ann}}(dH_{\rm c2}^{||c}/dT)$	& $-\mu_0dH_{\rm c2}^{||ab}/dT$ 	& $\delta_{t_{\rm ann}}(dH_{\rm c2}^{||ab}/dT)$		\\
							& (K)			&								& (T/K)						& 									& (T/K)						& 										\\\hline\hline
B$_{488}[0~{\rm h}]$			& 31.54(5)	&								& 1.58(3)						&									& 4.6(1)						& 										\\
B$_{488}[3~{\rm h}]$			& 33.07(5)	& +4.9\%							& 1.59(2)						& +0.6\%								& 5.8(1)						& +26.1\%									\\
\hline\hline		
\end{tabular}
\end{table*}
\begin{table*}[t!]
\caption{Evolution of $H_{\rm c2}^{||c}(0)$, $H_{\rm c2}^{||ab}(0)$, and $\gamma_H$ with annealing time $t_{\rm ann}$ for samples B$_{488}[0~{\rm h}]$ and B$_{488}[3~{\rm h}]$. The changes with annealing $\delta_{t_{\rm ann}}(H_{\rm c2}^{||c}(0))$, $\delta_{t_{\rm ann}}(H_{\rm c2}^{||ab}(0))$, and $\delta_{t_{\rm ann}}(\gamma_H)$ were calculated applying Eq.~(\ref{eq2}).}
\label{table4}
\begin{tabular}{ p{30mm} c c c c c c c c c c c c}
\hline\hline	
Sample						& $\mu_0H_{\rm c2}^{||c}(0)$ 	& $\delta_{t_{\rm ann}}(H_{\rm c2}^{||c}(0))$	& $\mu_0H_{\rm c2}^{||ab}(0)$	& $\delta_{t_{\rm ann}}(H_{\rm c2}^{||ab}(0))$		& $\gamma_H$	& $\delta_{t_{\rm ann}}(\gamma_H)$	\\
							& (T)						& 									& (T)						& 										& 				& 								\\\hline\hline
B$_{488}[0~{\rm h}]$			& 34.6(7)					& 									& 101(3)					&										& 2.9(2)			&								\\
B$_{488}[3~{\rm h}]$			& 36.5(5)					& +5.5\%								& 133(3)					& +31.7\%									& 3.6(2)			& +24.1\%							\\	
\hline\hline	
\end{tabular}
\end{table*}
Field dependent magnetization measurements were performed to further investigate the superconducting properties of the samples A$_{413}[t_{\rm ann}]$, A$_{488}[t_{\rm ann}]$, A$_{563}[t_{\rm ann}]$, and A$^*_{488}[t_{\rm ann}]$. In Fig.~\ref{fig4} the corresponding zfc magnetization curves measured at $T=2.0$~K with variable $t_{\rm ann}$ are presented. The internal magnetic field $H_{\rm int}$ was calculated by correcting the applied magnetic field $H$ for the demagnetization of the samples 
\begin{equation}
\label{eq3}
H_{\rm int}=H-DM,
\end{equation}
where $D$ is the demagnetization factor. The dimensions of the crystals used in this experiment were \mbox{$\sim2\times2\times0.5$~mm$^3$}, yielding $D\simeq0.8$ for the measurements with $H$ applied along the $c$-axis being the shortest dimension.\cite{Osborn1945} Hence, it was possible to determine the magnetization $M$ as a function of $H_{\rm int}$. In Fig.~\ref{fig4}(a) the $M(H_{\rm int})$ data for $t_{\rm ann}=0$~h are presened. All samples show rather poor superconducting properties. Although $M(H_{\rm int})\simeq -H_{\rm int}$ for low magnetic fields (almost ideal diamagnetism), the $M(H_{\rm int})$ curves strongly deviate from this linear behavior for field exceeding $1-2$~mT, indicating a rather small out-of-plane lower critical field $H_{\rm c1}^{||c}$. By means of the relation between $H_{\rm c1}^{||c}$ and the in-plane magnetic penetration depth $\lambda_{ab}$:
\begin{equation}
\label{eq4}
\mu_0H_{\rm c1}^{||c}=\frac{\Phi_0}{4\pi\lambda_{ab}^2}\left(\ln\kappa_{ab}+\frac{1}{2}\right),
\end{equation}
it was argued that a very small $\mu_0H_{\rm c1}\simeq0.3$~mT is consistent with a large $\lambda\simeq1-2$~$\mu$m.\cite{Bosma2012} However, this behavior is drastically changed with annealing as seen in Figs.~\ref{fig4}(b) and \ref{fig4}(c). Although the measurements for sample A$_{413}[t_{\rm ann}]$ reveal no obvious change with increasing $t_{\rm ann}$, the samples A$_{488}[t_{\rm ann}]$ and A$^*_{488}[t_{\rm ann}]$ show both a considerably higher diamagnetic response at higher $H_{\rm int}$, indicating an improved screening of the applied magnetic field. By defining $H_{\rm c1}$ as the magnetic field where the curves deviate from ideal diamagnetism, the best sample A$^*_{488}[t_{\rm ann}]$ yields a considerably larger $\mu_0H_{\rm c1}\simeq10$~mT compared to the estimate $\lesssim1$~mT for the as-grown samples. Such a large value of 10~mT is consistent with $\lambda\simeq270$~nm, assuming a realistic Ginzburg-Landau parameter $\kappa_{ab}\simeq100$ in Eq.~(\ref{eq4}). For a quantitative comparison of the superconducting properties of the different samples the superconducting susceptibility $\chi_{\rm sc}(\mu_0H_{\rm int})$ was estimated using the relation:
\begin{equation}
\label{eq5}
\chi_{\rm sc}(\mu_0H_{\rm int})=\frac{M(\mu_0H_{\rm int})}{H_{\rm int}}.
\end{equation}
In Table~\ref{table2} $\chi_{\rm sc}(1~{\rm mT})$ and $\chi_{\rm sc}(10~{\rm mT})$ are listed. Comparing $\chi_{\rm sc}(\mu_0H_{\rm int})$ for the sample A$_{413}[t_{\rm ann}]$ with increasing $t_{\rm ann}$, no improvement of the diamagnetic response was found with annealing. However, for all other samples A$_{488}[t_{\rm ann}]$, A$_{563}[t_{\rm ann}]$, and A$^*_{488}[t_{\rm ann}]$ both, $\chi_{\rm sc}(1~{\rm mT})$ and $\chi_{\rm sc}(10~{\rm mT})$ increase substantially with increasing $t_{\rm ann}$. Whereas the improvement of screening in 10~mT indicates an increase of critical current density, the changes observed in very low magnetic fields are rather related to an increase of $H_{\rm c1}$ connected with a decrease of $\lambda$. This suggests that the changes induced by annealing directly influence the density and the mobility of the charge carriers in the superconducting phase.
\\\indent
Besides magnetization, also resistivity experiments are expected to exhibit pronounced changes with annealing. Resistivity studies may provide independent and complementary information to the magnetization experiments. Whereas magnetization measurements probe the global macroscopic properties of a sample, its resistivity is sensitive to microscopic currents flowing through this mesoscopic phase separated material. Figure~\ref{fig5} shows the in-plane resistivity $\rho$ for the Rb$_x$Fe$_{2-y}$Se$_2$ single crystal, measured in zero magnetic field by cooling from 300 to 5~K. The measurements were performed on the as-grown sample (B$_{488}[0~{\rm h}]$) and were repeated after annealing in $488$~K for 3~h (B$_{488}[3~{\rm h}]$) using the same contacts. A clear reduction of $\rho$ in the normal state was found together with an increase of $T_{\rm c}$ from 31.5~K in the pristine sample to 33.1~K for the annealed sample (see Table~\ref{table3}), in very good agreement with the increase observed by magnetization (see Table~\ref{table1}). The hump in $\rho(T)$ between 100 and 150~K for the as-grown sample B$_{488}[0~{\rm h}]$ seen in Fig.~\ref{fig5} was earlier interpreted as a possible metal-insulator transition.\cite{Han2012} Such a transition would be likely related to the mesoscopic phase separation present in Rb$_x$Fe$_{2-y}$Se$_2$. In this picture the minority phase is connected with percolative paths along which electrical current may flow.\cite{Shen2011} Interestingly, this hump is strongly decreased with annealing at 488~K for 3~h, indicating that normal-state electric conductivity is enhanced in the annealed sample.
\\\indent
In Fig.~\ref{fig6}(a)-(d) the resistivity measurements at low temperatures performed on the pristine and annealed Rb$_x$Fe$_{2-y}$Se$_2$ single crystal B$_{488}[t_{\rm ann}]$ for various magnetic fields applied along the $c$-axis and in the $ab$-plane are presented. The transition temperature $T_{\rm c}$ is reduced with increasing $H$ for all configurations. In order to quantify this phase transition, the upper critical field $H_{\rm c2}$ is determined by following field and temperature at which 50\% of the normal state resistivity is suppressed [dashed line in Fig.~\ref{fig6}, panels (a)-(d)]. Figure~\ref{fig6}(e) shows the estimated upper critical field along the $c$-axis [$H_{\rm c2}^{||c}(T)$] and in the $ab$-plane [$H_{\rm c2}^{||ab}(T)$], for the as-grown and annealed sample. An increase of $T_{\rm c}(H)$ with annealing is observed in the whole temperature-field phase diagram. The slopes $-\mu_0dH_{\rm c2}^{||\alpha}/dT$ ($\alpha=c,ab$) of the phase boundaries for sufficiently high $H$ are summarized in Table~\ref{table3}. From these the upper critical fields at zero temperature were estimated applying the Werthamer-Helfand-Hohenberg (WHH) approximation\cite{Werthamer}
\begin{equation}
\label{eq6}
H_{\rm c2}(0)=-0.69\cdot T_{\rm c}\frac{dH_{\rm c2}}{dT},
\end{equation}
where $-dH_{\rm c2}/dT$ is defined as the maximal slope of the $H_{\rm c2}(T)$ curve in the vicinity of $T_{\rm c}$. Here we considered the linear part of the curve well below but not too far from $T_{\rm c}$, emphasized in Fig.~\ref{fig6}(e), which yields a more reliable estimate for the upper critical field of superconductors with an upturn curvature close to  $T_{\rm c}$.\cite{Bukowski2009} Interestingly, the upper critical field anisotropy 
\begin{equation}
\label{eq7}
\gamma_H=H_{\rm c2}^{||ab}/H_{\rm c2}^{||c},
\end{equation}
increases with annealing by 24.1\% (see Table~\ref{table4}). This suggests that thermally treated iron-chalcogenide superconductors with improved macroscopic physical properties are more anisotropic.
\\\indent
Besides investigating the properties in the superconducting state, it is also important to monitor the changes in normal-state properties of the Rb$_x$Fe$_{2-y}$Se$_2$ single crystals as a result of post annealing. In Fig.~\ref{fig7} we present the magnetic moment $m$ measured in 1~T and in 3~T for  A$_{413}[t_{\rm ann}]$, A$_{488}[t_{\rm ann}]$, A$_{563}[t_{\rm ann}]$, and A$^*_{488}[t_{\rm ann}]$ with $t_{\rm ann}=0$, 3, and 36~h. The magnetic moment in the normal state, recorded between 50 and 370~K, systematically increases with $t_{\rm ann}$ for {\it all} investigated samples. In the normal state the major component of the magnetic moment is stemming from the antiferromagnetic phase. However, some small ferromagnetic contribution is present in all Rb$_x$Fe$_{2-y}$Se$_2$ crystals, most likely due to a ferromagnetic impurity phase. From the measurements presented in Fig.~\ref{fig7} we determined the antiferromagnetic susceptibility $\chi_{\rm AFM}(T)$ according to
\begin{equation}
\label{eq8}
\chi_{\rm AFM}(T)=\frac{1}{\mathcal{M}}\cdot\frac{m(\mu_0H)-m(\mu_0H')}{H-H'},
\end{equation}
where $\mathcal{M}$ denotes the sample mass. Here, $\mu_0H$ and $\mu_0H'$ are 1 and 3~T, respectively. The ferromagnetic contribution to the magnetization is assumed to be constant in field and is derived accordingly
\begin{equation}
\label{eq9}
M_{\rm FM}(T)=\frac{m(\mu_0H)}{\mathcal{M}}-\chi_{\rm AFM}(T)\cdot H.
\end{equation}
The antiferromagnetic susceptibility for all the as-grown samples and those annealed for 3~h and for 36~h are shown in Fig.~\ref{fig8}(a). The ferromagnetic component of the magnetization $M_{\rm FM}(T)$ is shown in Fig.~\ref{fig8}(b). Sample  A$_{413}[t_{\rm ann}]$ remains unaffected by annealing, as already observed in the zfc magnetization experiments performed in the superconducting state as discussed above. However, for the samples A$_{488}[t_{\rm ann}]$, A$_{563}[t_{\rm ann}]$, and A$^*_{488}[t_{\rm ann}]$ the high-field susceptibility $\chi_{\rm AFM}(T)$ increases substantially with increasing $t_{\rm ann}$. In Table~\ref{table5} we list the observed values for  $\chi_{\rm AFM}(50~{\rm K})$ for all samples and $t_{\rm ann}$. Obviously, the change in $\chi_{\rm AFM}(50~{\rm K})$ is most pronounced for the sample A$_{563}[t_{\rm ann}]$, annealed at 563~K. In addition, the ferromagnetic component $M_{\rm FM}(T)$ is almost unchanged for sample A$_{413}[t_{\rm ann}]$, but increases for the samples A$_{488}[t_{\rm ann}]$, A$_{563}[t_{\rm ann}]$, and A$^*_{488}[t_{\rm ann}]$ with increasing $t_{\rm ann}$. Again, the change in $M_{\rm FM}(50~{\rm K})$ is maximal for sample A$_{563}[t_{\rm ann}]$.
\\\indent
\begin{table*}[t!]
\caption{Evolution of $\chi_{\rm AFM}(50~{\rm K})$ [see Eq.~(\ref{eq8})] and $M_{\rm FM}(50~{\rm K})$ [see Eq.~(\ref{eq9})] with annealing time $t_{\rm ann}$ for the samples A$_{413}[t_{\rm ann}]$, A$_{488}[t_{\rm ann}]$, A$_{563}[t_{\rm ann}]$, and A$^*_{488}[t_{\rm ann}]$. The changes with annealing $\delta_{t_{\rm ann}}(\chi_{\rm AFM}(50~{\rm K}))$ and $\delta_{t_{\rm ann}}(M_{\rm FM}(50~{\rm K}))$ were calculated applying Eq.~(\ref{eq2}).}
\label{table5}
\begin{tabular}{p{25mm}  c c c c }
\hline\hline
Sample							& $\chi_{\rm AFM}(50~{\rm K})$	& $\delta_{t_{\rm ann}}(\chi_{\rm AFM}(50~{\rm K}))$ 	& $M_{\rm FM}(50~{\rm K})$		& $\delta_{t_{\rm ann}}(M_{\rm FM}(50~{\rm K}))$		\\
								& $(10^{-8}~{\rm m}^3/{\rm kg})$	& 											& $(10^{-3}~{\rm Am}^2/{\rm kg})$	& 											\\\hline\hline
A$_{413}[0~{\rm h}]$				& 2.020(1)					&											& 2.79(1)						&											\\
A$_{413}[3~{\rm h}]$				& 2.006(1)					& $-$0.7\%									& 2.75(1)						& $-$1.4\%									\\
A$_{413}[36~{\rm h}]$				& 2.044(1)					& +1.2\%										& 2.79(1)						& $\pm0.0$\%									\\\hline
A$_{488}[0~{\rm h}]$				& 1.808(1)					&											& 2.77(1)						& 											\\
A$_{488}[3~{\rm h}]$				& 1.883(1)					& +4.1\%										& 5.91(1)						& +113\%										\\
A$_{488}[36~{\rm h}]$				& 1.953(1)					& +8.0\%										& 10.73(1)					& +287\%										\\\hline
A$_{563}[0~{\rm h}]$				& 2.400(1)					&											& 6.91(1)						&											\\
A$_{563}[3~{\rm h}]$				& 2.598(1)					& +8.3\%										& 17.25(1)					& +150\%										\\
A$_{563}[36~{\rm h}]$				& 2.778(1)					& +15.8\%										& 28.78(1)					& +317\%										\\\hline
A$^*_{488}[0~{\rm h}]$				& 1.796(1)					&											& 2.67(1)						&											\\
A$^*_{488}[3~{\rm h}]$				& 1.883(1)					& +4.8\%										& 3.43(1)						& +28.5\%										\\
A$^*_{488}[36~{\rm h}]$ 				& 1.947(1)					& +8.4\%										& 15.92(1)					& +496\%										\\\hline\hline
\end{tabular}
\end{table*}
The effect of annealing on the magnetic and superconducting properties of Rb$_x$Fe$_{2-y}$Se$_2$ single crystals was further investigated by means of transverse-field (TF) and zero-field (ZF) $\mu$SR experiments. The $\mu$SR measurements are based on the observation of the time evolution of the muon spin polarization. (For a detailed description of the $\mu$SR technique see \emph{e.g.} Ref.~\cite{Amato}.) For these experiments two mosaics of samples were prepared: \mbox{(i) C$_{488}[0~{\rm h}]$}, consisting of three as-grown Rb$_x$Fe$_{2-y}$Se$_2$ single crystals, and (ii) C$_{488}[60~{\rm h}]$, consisting of three Rb$_x$Fe$_{2-y}$Se$_2$ single crystals simultaneously annealed in 488~K for 60~h. Previous $\mu$SR experiments revealed that Rb$_x$Fe$_{2-y}$Se$_2$ consists of a magnetic ($\sim90\%$) and a non-magnetic superconducting ($\sim10\%$) phase.\cite{Shermadini2012} In order to investigate the superconducting properties, a field of 70~mT was applied transverse to the initial muon spin polarization and parallel to the crystallographic $c$-axis. In this TF configuration the muons probe the local magnetic field distribution $P(B)$ of the vortex lattice formed in the superconducting areas. Simultaneously, the signal stemming from the magnetic regions of the sample is suppressed, since the superposition of the strong internal field and the weak external field leads to a fast depolarization and to a loss of asymmetry.
\\\indent
\begin{figure}[t!]
\centering
\includegraphics[width=\linewidth]{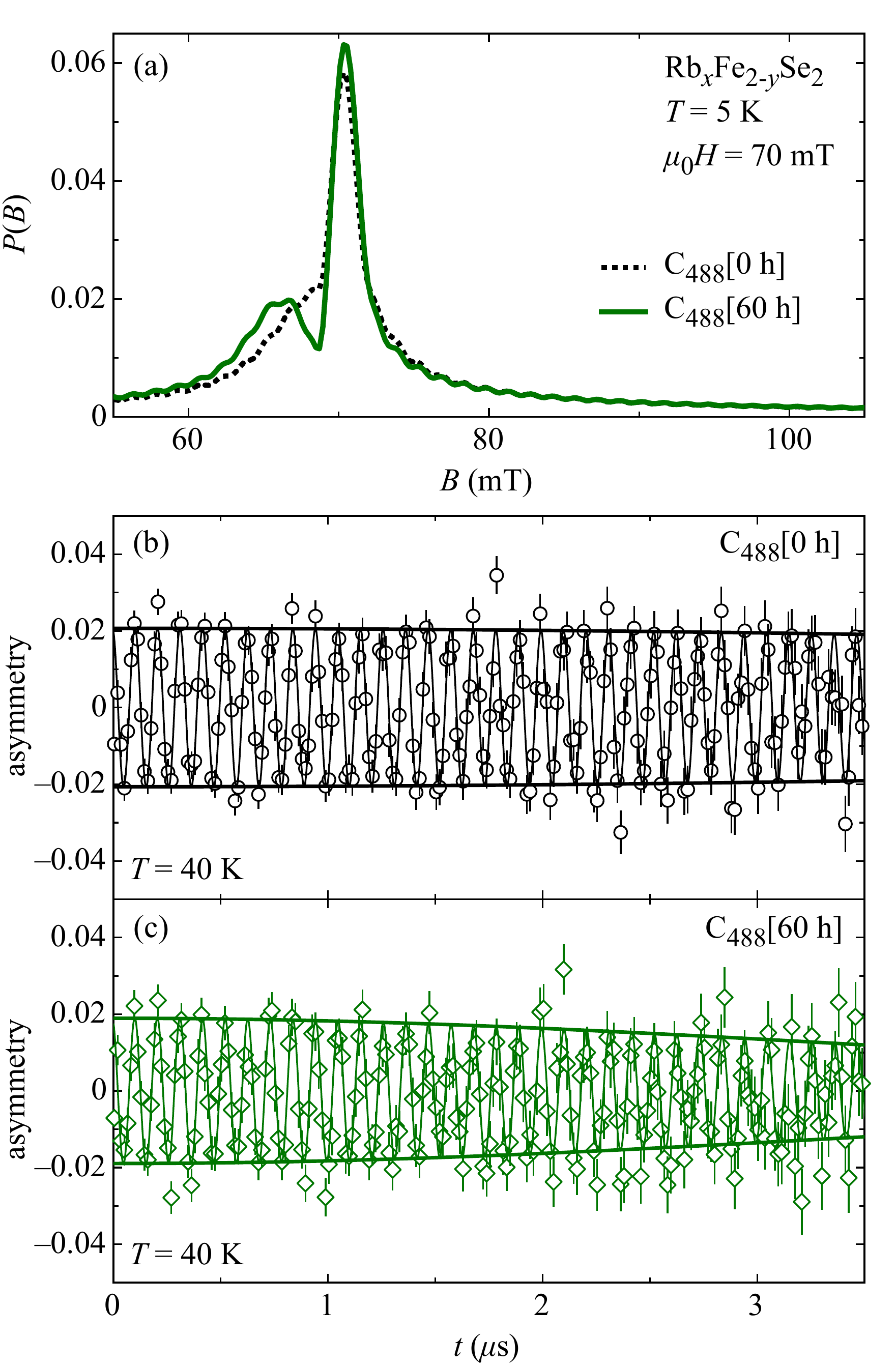}
\caption{(color online) Results of the TF $\mu$SR investigation of as-grown and 60~h annealed  Rb$_x$Fe$_{2-y}$Se$_2$ single crystals, C$_{488}[0~{\rm h}]$ and C$_{488}[60~{\rm h}]$, in an magnetic field of 70~mT applied along the $c$-axis. (a) $P(B)$ for both samples at 5~K. The line shape for C$_{488}[60~{\rm h}]$ is more asymmetric compared to that for the as-grown sample C$_{488}[0~{\rm h}]$. (b) and (c) $\mu$SR time spectra at 40~K for sample C$_{488}[0~{\rm h}]$ and C$_{488}[60~{\rm h}]$. The thin solid line is a fit to the data assuming a single relaxation rate $\sigma$. The thick solid line is the envelope of the oscillating function. The data for the annealed sample C$_{488}[60~{\rm h}]$ exhibit a significantly faster damping.}
\label{fig9}
\end{figure}
\begin{figure}[t!]
\centering
\includegraphics[width=\linewidth]{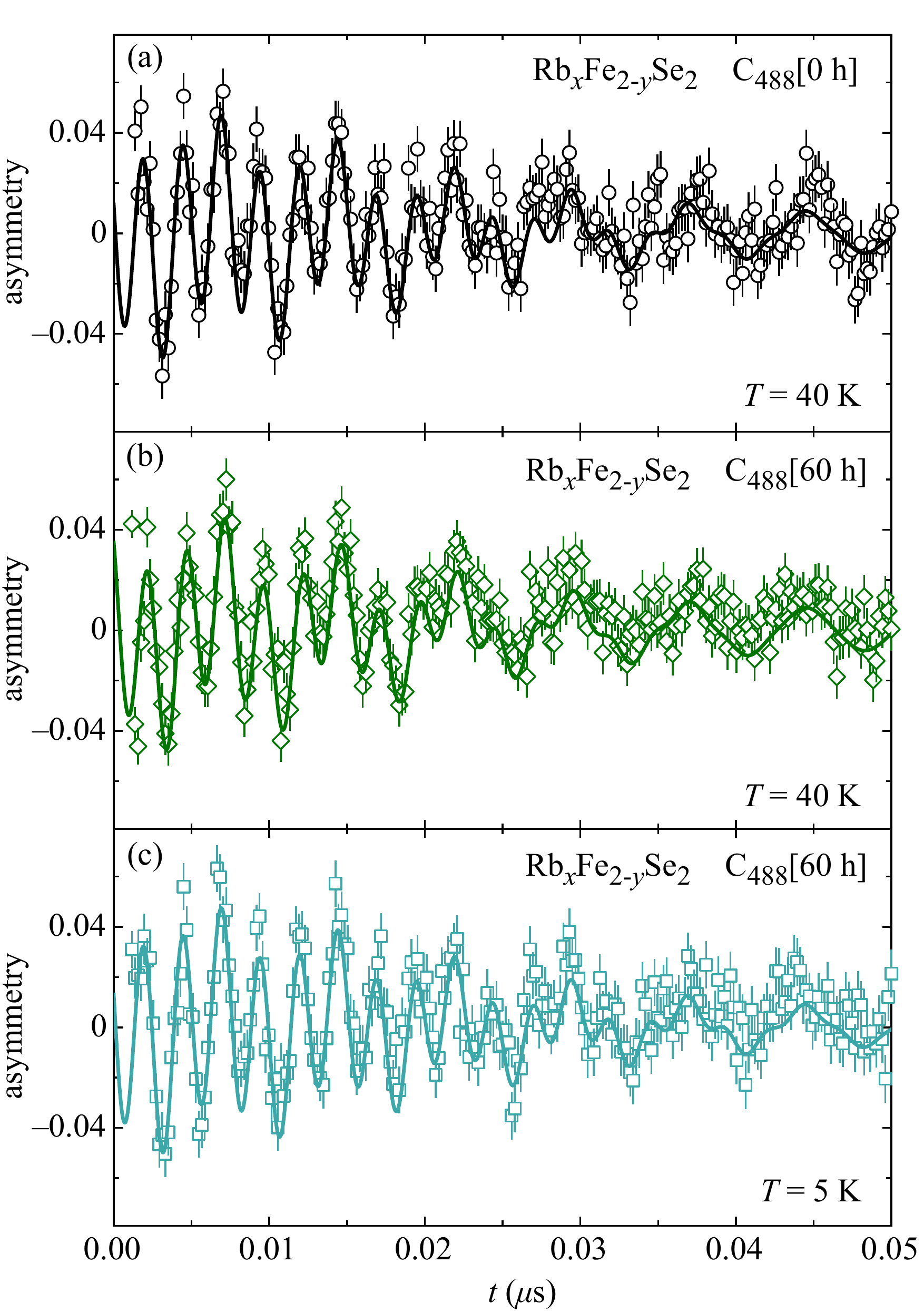}
\caption{(color online) Results of the ZF $\mu$SR investigation of as-grown and 60~h annealed Rb$_x$Fe$_{2-y}$Se$_2$ single crystals, C$_{488}[0~{\rm h}]$ and C$_{488}[60~{\rm h}]$. All data were modeled assuming two internal magnetic fields $B_{\rm int,1}\approx 1$~T and $B_{\rm int,2}\approx 3$~T.}
\label{fig10}
\end{figure}
Consistent with the above presented macroscopic magnetization and resistivity results, also the intrinsic superconducting properties are significantly improved after annealing. The lineshape of the local magnetic field distribution $P(B)$ of C$_{488}[60~{\rm h}]$ shown in Fig.~\ref{fig9}(a) is more asymmetric as compared to that of C$_{488}[0~{\rm h}]$, indicating the presence of a more homogeneous and more regular vortex lattice in the superconducting regions. Note that the sharp peak of $P(B)$ at 70~mT is stemming from the signal of background muons, whose spins rotate simply in the applied magnetic field. A more detailed analysis of the as obtained $P(B)$ yields that the shielding of the magnetic field for C$_{488}[60~{\rm h}]$ is substantially larger due to a reduction of the first moment $<B>$ of $P(B)$ by $\sim5\%$. This is surprising, since the microscopic in-plane magnetic penetration depth $\lambda_{ab}(0)\simeq258(2)$~nm,\cite{Shermadini2012} as well as the total asymmetry of the superconducting part remain essentially unchanged after 60~h annealing [see Fig.~\ref{fig9}(b)]. These results imply that the volume fraction of the magnetic and the non-magnetic phase is unaffected by annealing, in contradiction to the conclusions of a neutron diffraction study, reporting a reduction of the minority phase after annealing of Rb$_x$Fe$_{2-y}$Se$_2$ single crystals for 100~h at 488~K.\cite{Pomjakushin2012} This discrepancy might arise from the difference in $T_{\rm p}$ of the samples studied here (489~K) and in Ref.~\onlinecite{Pomjakushin2012} (475~K).
\\\indent
\begin{figure*}[t!]
\centering
\includegraphics[width=\linewidth]{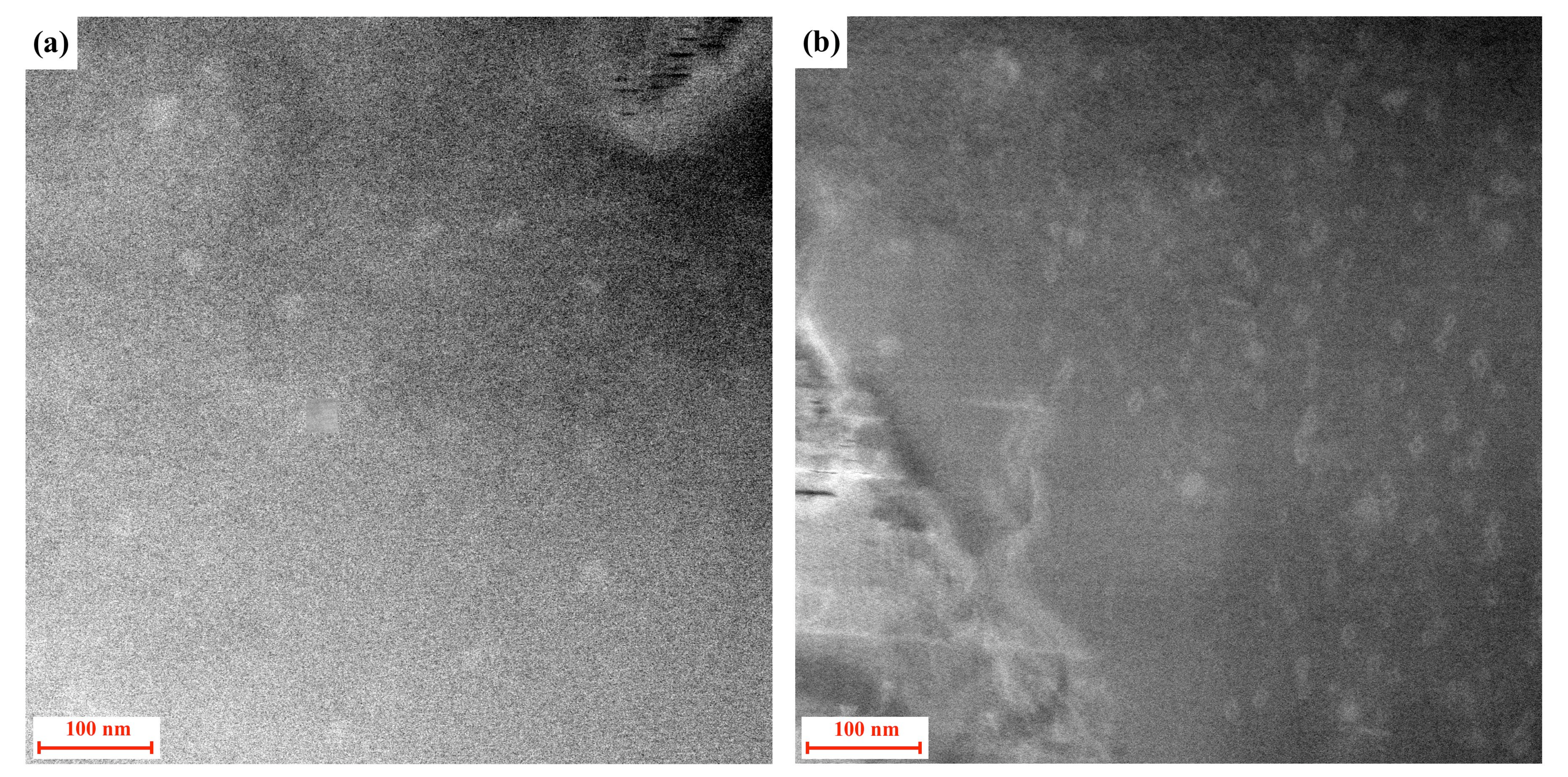}
\caption{(color online) STEM images of as-grown and annealed Rb$_x$Fe$_{2-y}$Se$_2$ single crystals D$_{488}[0~{\rm h}]$ and D$_{488}[3~{\rm h}]$. The microstructure caused by mesoscopic phase separation in the annealed sample D$_{488}[3~{\rm h}]$, shown in panel (b), is modified compared to the one of the as-grown sample D$_{488}[0~{\rm h}]$, shown in panel (a).}
\label{fig11}
\end{figure*}
Importantly, the normal-state relaxation rate $\sigma$ of the $\mu$SR time spectra derived from the data at 40~K (well above $T_{\rm c}$) increases drastically with $t_{\rm ann}$ [see Fig.~\ref{fig9}(b) and (c)]. Whereas for the as-grown sample C$_{488}[0~{\rm h}]$ $\sigma=0.141(33)~\mu$s$^{-1}$, the relaxation rate of the 60~h annealed sample C$_{488}[60~{\rm h}]$ is considerably larger ($\sigma=0.303(43)~\mu$s$^{-1}$). This indicates a substantially increased field inhomogeneity in the non-magnetic part of the sample. Since the volume fraction is unchanged during annealing, this suggests that the microstructure of the sample caused by mesoscopic phase separation is modified by annealing at 488~K, in such a way that the individual size of the non-magnetic regions is reduced and their number is increased, {\it but their total volume remains unaffected}. 
\\\indent
In order to examine our samples for the internal magnetic field distribution when no magnetic field is applied, low temperature ZF $\mu$SR experiments were performed on the same Rb$_x$Fe$_{2-y}$Se$_2$ single crystals. Consistent with the results of the TF experiments, the total volume of the non-magnetic regions was found to be $\sim10\%$ of the total sample volume only. In the ZF data a clear oscillating signal may be found in all samples for very short time scales as shown in Fig.~\ref{fig10}. An analysis of the time evolution of this signal revealed that two internal magnetic fields $B_{\rm int,1}\approx1$~T and $B_{\rm int,2}\approx3$~T are present in the samples. In analogy to the evolution of the magnetic volume fraction, $B_{\rm int,1}$ and $B_{\rm int,2}$ are not affected by annealing at 488~K. They are directly proportional to the iron moment in the antiferromagnetic phase. Moreover, annealing again does not affect the ratio $B_{\rm int,1}/B_{\rm int,2}$. Hence, no changes in the internal magnetic fields were observed by $\mu$SR after annealing the as-grown Rb$_x$Fe$_{2-y}$Se$_2$ single crystals, even though the macroscopic superconducting properties were substantially improved (see Figs.~\ref{fig3} and \ref{fig4}).
\\\indent
In order to visualize microscopic changes in the phase separation of our Rb$_x$Fe$_{2-y}$Se$_2$ single crystals with annealing, additional STEM images were taken on as-grown and annealed samples D$_{488}[0~{\rm h}]$ and D$_{488}[3~{\rm h}]$ (see Fig.~\ref{fig11}). The microstructure caused by mesoscopic phase separation in the annealed sample D$_{488}[3~{\rm h}]$, shown in Fig.~\ref{fig11}(b), is modified compared to the one of the as-grown sample D$_{488}[0~{\rm h}]$, shown in Fig.~\ref{fig11}(a). Whereas a few inclusions of the minority phases only are observed at the surface of D$_{488}[0~{\rm h}]$, sample D$_{488}[3~{\rm h}]$ reveals plenty of such inclusions in the same area. However, the inclusions of the minority phase of sample D$_{488}[3~{\rm h}]$ are in general smaller in size than the ones of the as-grown sample D$_{488}[0~{\rm h}]$, in agreement with the results of the above $\mu$SR experiments. 
\\\indent
\section{Discussion}
The superconducting and normal-state properties of mesoscopically phase separated Rb$_x$Fe$_{2-y}$Se$_2$, where non-magnetic regions exist in a magnetic surrounding, are strikingly similar to those expected for granular superconductors. From early work on granular superconductors it is known that the macroscopic properties of such materials studied by various techniques may vary substantially, depending on the particular grain-size distribution and their coupling by Josephson links.\cite{Ambegaokar1963, Ebner1985, Clem1988}. Importantly, granular superconductors may easily appear as bulk superconducting, however, their superconducting-state parameters, such as the magnetic penetration depth $\lambda$, the coherence length $\xi$, and the lower and upper critical fields ($H_{\rm c1}$ and $H_{\rm c2}$) differ substantially from those of related non-granular superconductors. Such a scenario may also hold for mesoscopically phase separated Rb$_x$Fe$_{2-y}$Se$_2$, since various experimental techniques provide quite different values for $\lambda$. For Rb$_x$Fe$_{2-y}$Se$_2$ recent $\mu$SR studies yielded $\lambda_{ab}(0)\simeq250-260$~nm,\cite{Shermadini2012, Charnukha2012_b} in agreement with $\lambda_{ab}(0)\simeq290$~nm obtained for K$_x$Fe$_{2-y}$Se$_2$ by means of high field nuclear magnetic resonance (NMR) experiments.\cite{Torchetti2011} These values are considerably smaller than those usually obtained by macroscopic techniques ($\lambda_{ab}(0)\simeq1.6-2.2~\mu$m).\cite{Bosma2012, Charnukha2012, Homes2012} In a mesoscopically phase separated superconductor macroscopic experiments yield an effective magnetic penetration depth which is a measure of the length scale over which the magnetic field penetrates the sample. On the other hand, $\mu$SR is a microscopic probe of the vortex state and is only sensitive to the superconducting fraction of the sample. Therefore, $\mu$SR measures a value of the magnetic penetration depth which is closer to the instrinsic value than the values usually obtained by macroscopic techniques. Since so far no single-phase superconducting $A_x$Fe$_{2-y}$Se$_2$ sample could be synthesized, it should not be excluded that granularity might be an important ingredient for the appearance of superconductivity in this system.
\\\indent
\begin{figure}[t!]
\centering
\includegraphics[width=\linewidth]{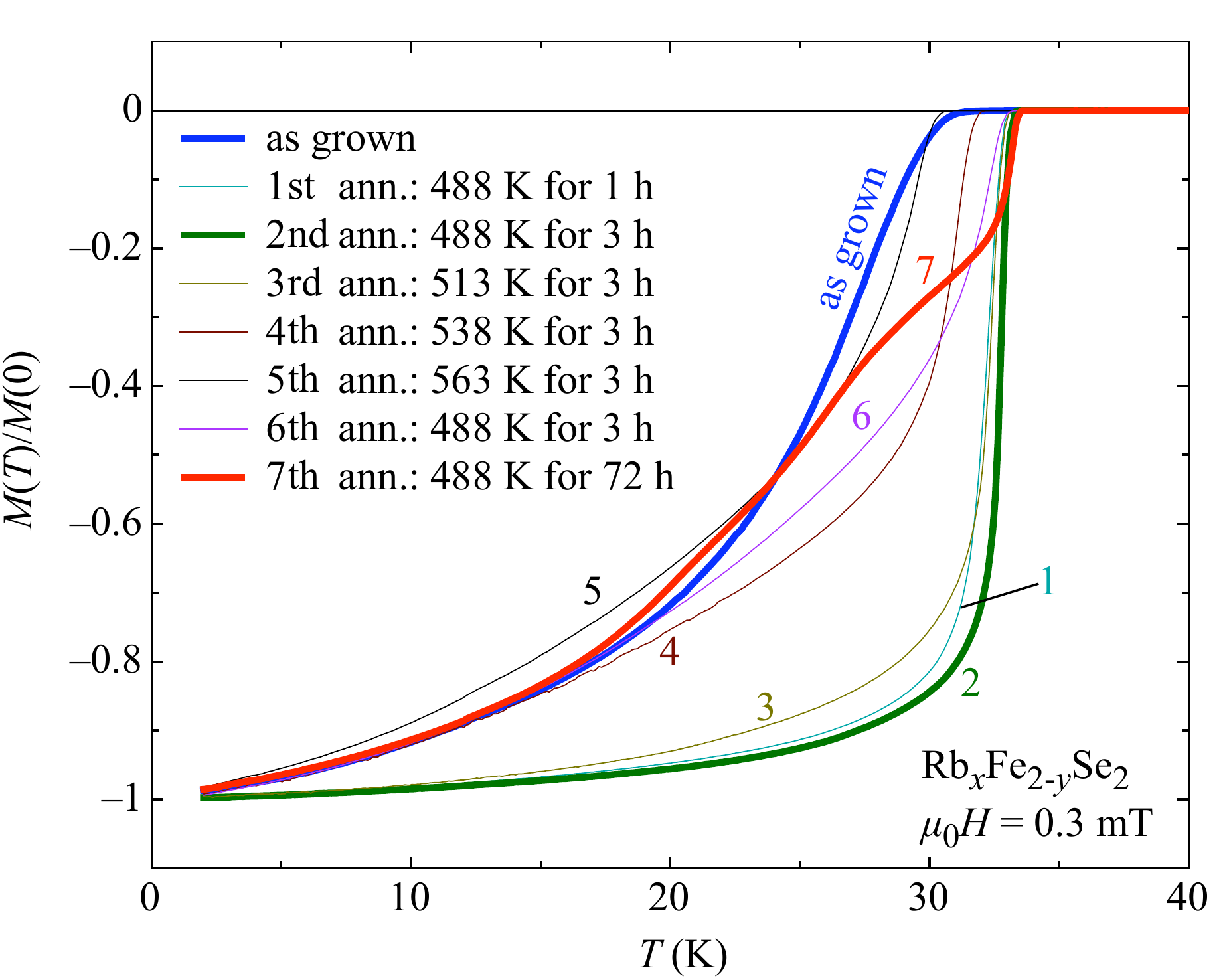}
\caption{(color online) Series of temperature dependent zfc magnetization measurements on a Rb$_x$Fe$_{2-y}$Se$_2$ single crystal in 0.3~mT. The curves obtained after the various post-annealings of the sample are labeled by the respective number.}
\label{fig12}
\end{figure}
As strongly suggested by the presented magnetization and resistivity data, pronounced changes of the physical properties of Rb$_x$Fe$_{2-y}$Se$_2$ are caused by tuning the annealing conditions. Whereas annealing at 413~K, well below $T_{\rm p}$, does not lead to any significant change in magnetic and transport properties, annealing just at $T_{\rm p}$, the onset of phase separation, favors the enhancement of superconductivity. Accordingly, $T_{\rm c}$ increases, the transition sharpens, the normal-state resistivity decreases, and $H_{\rm c2}$ increases. However, after annealing at 563~K, well above $T_{\rm p}$, all superconducting properties get drastically suppressed. In addition, the antiferromagnetic susceptibility and the ferromagnetic saturation magnetization of the investigated samples systematically increase. This may be related to the change in iron valency as observed in annealed K$_{0.8}$Fe$_{1.6}$Se$_2$,\cite{Simonelli2012} or with an increase of Fe-based impurity phases. 
\\\indent
A recent neutron diffraction study of the Rb$_x$Fe$_{2-y}$Se$_2$ system reports a pronounced reduction of the 122 minority phase when the samples were annealed at 488~K for 100~h.\cite{Pomjakushin2012} However, the present $\mu$SR experiments yield clear evidence that the volume fraction of the two phases remains unchanged by annealing, while the field inhomogeneity in the non-magnetic parts of the sample increases substantially. This implies that the microstructure caused by mesoscopic phase separation in the sample is modified by annealing just at $T_{\rm p}$ in such a way that the size of non-magnetic regions is reduced, and the number of regions is increased, but their total volume remains unaffected. Since the $\mu$SR results clearly demonstrate that the total volume of the minority phase is constant, even after 60~h of annealing, this rearrangement of the coexisting phases leads to the conclusion that changes of the coupling between these regions must be related to the improvement of superconductive properties. Whereas, 488~K was chosen to match the onset of phase separation $T_{\rm p}\simeq489$~K in the single crystals studied here, the samples used in the neutron diffraction study had a significantly lower $T_{\rm p}\simeq475$~K.\cite{Pomjakushin2012} Therefore, the observed reduction of the minority phase found by the neutron study might be due to a partial degradation of the minority phase as a result of 100~h annealing at temperatures exceeding $T_{\rm p}$. That this scenario appears to be reasonable is further supported by the data presented in Fig.~\ref{fig12}, where a series of magnetization measurements are shown for a Rb$_x$Fe$_{2-y}$Se$_2$ single crystal of a similar batch as the one used above. Here, always the same temperature dependence of the zfc magnetization measurement in a magnetic field of $\mu_0H=0.3$~mT along the $c$-axis was performed after each subsequent annealing of the sealed single crystal in a quartz ampoule. Note that $T_{\rm c}$ of the as-grown sample is easily shifted to higher values by an annealing at 488~K for some hours. However, after the subsequent annealings during which the temperature was modestly increased up to 563~K, superconductivity is strongly suppressed as seen by the decrease of $T_{\rm c}$ and the broadening of the transition. During the final annealing, again the optimal annealing temperature of 488~K was chosen, this time for a very long annealing time up to 72~h. However, superconductivity did not fully recover. Obviously, the short annealings at temperatures exceeding $T_{\rm p}$ formed additional magnetic phases, which cannot be reversed anymore, even by choosing a very long annealing time.
\\\indent
All changes of superconducting and magnetic properties caused by annealing are evidently related to changes in the microstructure of the sample caused by mesoscopic phase separation in Rb$_x$Fe$_{2-y}$Se$_2$. The difference of the superconducting properties between the as-grown and annealed single crystals are likely explained by assuming that inhomogeneities (in particular phase boundaries and/or stripes) are necessary to enhance superconductivity.\cite{Bianconi1996, Moon2010, Das2011, Andersen2011, Wittlin2012} In the present case, the existing boundaries between the magnetic majority regions and non-magnetic minority regions may play the role of such inhomogeneities. In the current case, reviewing the changes observed of the superconducting and normal-state properties with annealing, it is likely that the intergrain coupling between magnetic and non-magnetic domains is crucial. Annealing of Rb$_x$Fe$_{2-y}$Se$_2$ single crystals just at $T_{\rm p}$ favors the mesoscopic phase separation in such a way that domain boundaries are further developed, improving all superconducting properties. However, if the samples are annealed at higher temperature, the superconducting phase degrades and by that it is more difficult to build up a percolative network favorable for superconductivity. In total $\sim10\%$ of the sample remains superconducting in a magnetic field of $70$~mT, whereas its macroscopic properties strongly depend on the optimal coupling between the superconducting regions, being strongly field and temperature dependent. This scenario appears similar to that of a granular superconductor in which the macroscopic physics are directly connected to the microscopic Josephson coupling between the individual grains. In addition, all changes in the phase separation may be related to changes in crystal structure and lattice parameters.\cite{Pomjakushin2012} Thus, internal pressure on the superconducting and non-superconducting domains may be likely involved in the appearance of superconductivity. Besides, also metallic nano-clusters were reported to show enhanced superconducting properties.\cite{Kresin2012} 
\section{Conclusions}
Extended magnetization and resistivity measurements of Rb$_{x}$Fe$_{2-y}$Se$_{2}$ single crystals revealed that post annealing at a temperature well below the onset temperature of phase separation  $T_{\rm p}$ neither changes the magnetic nor the superconducting properties of the crystals. Annealing at a temperature above $T_{\rm p}$ reduces the value of $T_{\rm c}$ drastically and suppresses antiferromagnetic order. However, annealing at 488~K, just at $T_{\rm p}$ leads to a substantial increase of $T_{\rm c}$ and sharpens the transition to the superconducting state. These results suggest that the superconducting properties of mesoscopically phase separated Rb$_{x}$Fe$_{2-y}$Se$_{2}$ can be tuned by the annealing temperature. In addition, $\mu$SR and STEM investigations indicate that non-magnetic regions of the sample rearrange with annealing at 488~K in such a way that their individual size is reduced and the number of regions is increased, but their total volume remains unaffected. At temperatures exceeding $T_{\rm p}$, where the majority $I4/m$ phase prevails, ferromagnetism is enhanced with annealing time, but is presumably detrimental to the formation of the superconducting phase. In conclusion, by annealing single crystals of Rb$_{x}$Fe$_{2-y}$Se$_{2}$ the microstructure of the crystals arising from mesoscopic phase separation is changed, leading to an improvement of the superconducting properties and an enhancement of $T_{\rm c}$.\\

\begin{acknowledgments}
Helpful discussions with K.~A.~M\"uller, V.~Yu.~Pomjakushin, and B.~M.~Wojek are gratefully acknowledged. S.~W.~thanks A.~Feierls and S.~R\"osch for their help during part of the experiments. The $\mu$SR measurements were performed at the Swiss Muon Source, Paul Scherrer Institute, Villigen, Switzerland. This work was supported by the Swiss National Science Foundation, the NCCR program MaNEP, the National Science Centre of Poland based on decision No.~DEC-2011/01/B/ST3/02374, and the European Regional Development Fund within the Innovative Economy Operational Programme 2007-2013 No~POIG.02.01-00-14-032/08.
\end{acknowledgments}


\begin{thebibliography}{99}



\bibitem{Hsu2009} F.~C.~Hsu, J.~Y.~Luo, K.-W.~Yeh, T.-K.~Chen, T.-W.~Huang, P.-M.~Wu, Y.-C.~Lee, Y.-L.~Huang, Y.-Y.~Chu, D.-C.~Yan, and M.-K.~Wu, Proc. Natl. Acad. Sci. USA {\bf 105}, 14262 (2008).

\bibitem{Pomjakushina2009} E.~Pomjakushina, K.~Conder, V.~Pomjakushin, M.~Bendele, and R.~Khasanov, Phys. Rev. B {\bf 80}, 024517 (2009).

\bibitem{Medvedev2009} S.~Medvedev, T.~M.~McQueen, I.~A.~Troyan, T.~Palasyuk, M.~I.~Eremets, R.~J.~Cava, S.~Naghavi, F.~Casper, V.~Ksenofontov, G.~Wortmann, and C.~Felser, Nat. Mater. {\bf 8}, 630 (2009).

\bibitem{Mizuguchi2009} Y.~Mizuguchi, F.~Tomioka, S.~Tsuda, T.~Yamaguchi, and Y.~Takano, J. Phys. Soc. Jpn. {\bf 78}, 074712 (2009).

\bibitem{Sales2009} B.~C.~Sales, A.~S.~Sefat, M.~A.~McGuire, R.~Y.~Jin, D.~Mandrus, and Y.~Mozharivskyj, Phys. Rev. B {\bf 79}, 094521 (2009).

\bibitem{Guo2010} J.~Guo, S.~Jin, G.~Wang, S.~Wang, K.~Zhu, T.~Zhou, M.~He, and X.~Chen, Phys. Rev. B {\bf 82}, 180520(R) (2010).

\bibitem{Krzton2011} A.~Krzton-Maziopa, Z.~Shermadini, E.~Pomjakushina, V.~Pomjakushin, M.~Bendele, A.~Amato, R.~Khasanov, H.~Luetkens, and K.~Conder, J. Phys.: Condens. Matter {\bf 23}, 052203 (2011).

\bibitem{Li2011} C.-H.~Li, B.~Shen, F.~Han, X.~Zhu, and H.-H.~Wen, Phys. Rev. B {\bf 83}, 184521 (2011).

\bibitem{Liu2011} R.~H.~Liu, X.~G.~Luo, M.~Zhang, A.~F.~Wang, J.~J.~Ying, X.~F.~Wang, Y~.~J.~Yan, Z.~J.~Xiang, P.~Cheng, G.~J.~Ye, Z.~Y.~Li, and X.~H.~Chen, Europhys. Lett. {\bf 94}, 27008 (2011).

\bibitem{Shermadini2011} Z.~Shermadini, A.~Krzton-Maziopa, M.~Bendele, R.~Khasanov, H.~Luetkens, K.~Conder, E.~Pomjakushina, S.~Weyeneth, V.~Pomjakushin, O.~Bossen, and A.~Amato, Phys. Rev. Lett. {\bf 106}, 117602 (2011).

\bibitem{Pomjakushin2011} V.~Yu.~Pomjakushin, E.~V.~Pomjakushina, A.~Krzton-Maziopa, K.~Conder, and Z.~Shermadini, J. Phys.: Condens. Matter {\bf 23}, 156003 (2011).

\bibitem{Bosma2012} S.~Bosma, R.~Puzniak, A.~Krzton-Maziopa, M.~Bendele, E.~Pomjakushina, K.~Conder, H.~Keller, and S.~Weyeneth, Phys. Rev. B {\bf 85}, 064509 (2012).

\bibitem{Tsurkan2011} V.~Tsurkan, J.~Deisenhofer, A.~G\"unther, H.-A.~Krug von Nidda, S.~Widmann, and A.~Loidl, Phys. Rev. B {\bf 84}, 144520 (2011).

\bibitem{Ying2011} J.~J.~Ying, X.~F.~Wang, X.~G.~Luo, A.~F.~Wang, M.~Zhang, Y.~J.~Yan, Z.~J.~Xiang, R.~H.~Liu, P.~Cheng, G.~J.~Ye, and X.~H.~Chen, Phys. Rev. B {\bf 83}, 212502 (2011).

\bibitem{Shermadini2012} Z.~Shermadini, H.~Luetkens, R.~Khasanov, A.~Krzton-Maziopa, K.~Conder, E.~Pomjakushina, H.-H.~Klauss, and A.~Amato, Phys. Rev. B {\bf 85}, 100501(R) (2012).

\bibitem{Pomjakushin2012} V.~Yu.~Pomjakushin, E.~V.~Pomjakushina, A.~Krzton-Maziopa, K.~Conder, D.~Chernyshov, V.~Svitlyk, and A.~Bosak, J. Phys.: Condens. Matter {\bf 24}, 435701(2012).

\bibitem{Shen2011} B.~Shen, B.~Zeng, G.~F.~Chen, J.~B.~He, D.~M.~Wang, H.~Yang, and H.~H.~Wen, Europhys. Lett. {\bf 96}, 37010 (2011).

\bibitem{Ricci2011} A.~Ricci, N.~Poccia, G.~Campi, B.~Joseph, G.~Arrighetti, L.~Barba, M.~Reynolds, M.~Burghammer, H.~Takeya, Y.~Mizuguchi, Y.~Takano, M.~Colapietro, N.~L.~Saini, and A.~Bianconi, Phys. Rev. B {\bf 84}, 060511(R) (2011).

\bibitem{Ricci2011_b} A.~Ricci, N.~Poccia, B.~Joseph, G.~Arrighetti, L.~Barba, J.~Plaisier, G.~Campi, Y.~Mizuguchi, H.~Takeya, Y.~Takano, N.~L.~Saini, and A.~Bianconi, Supercond. Sci. Technol. {\bf 24}, 082002 (2011).

\bibitem{Lei2011_c} H.~Lei and C.~Petrovic, Phys. Rev. B {\bf 83}, 184504 (2011).

\bibitem{Yan2011} Y.~J.~Yan, M~Zhang, A.~F.~Wang, J.~J.~Ying, Z.~Y.~Li, W.~Qin, X.~G.~Luo, J.~Q.~Li, J.~Hu, and X.~H.~Chen, Scientific Reports {\bf 2}, 212 (2012).

\bibitem{Wang2011} C.~N.~Wang, P.~Marsik, R.~Schuster, A.~Dubroka, M.~R\"ossle, C.~Niedermayer, G.~D.~Varma, A.~F.~Wang, X.~H.~Chen, T.~Wolf, and C.~Bernhard , Phys. Rev. B {\bf 85}, 214503 (2012) . 

\bibitem{Bosak2011} A.~Bosak, V.~Svitlyk, A.~Krzton-Maziopa, E.~Pomjakushina, K.~Conder, V.~Pomjakushin, A.~Popov, D.~de~Sanctis, D.~Chernyshov, (unpublished) arXiv/condmat:1112.2569v1 (2011).

\bibitem{Texier2012} Y.~Texier, J.~Deisenhofer, V.~Tsurkan, A.~Loidl, D.~S.~Isonov, G.~Friemel, and J.~Bobroff, 	Phys. Rev. Lett. {\bf 108}, 237002 (2012). 

\bibitem{Speller2012}  S.~C.~Speller, T.~B.~Britton, G.~M.~Hughes, A.~Krzton-Maziopa, E.~Pomjakushina, K.~Conder, A.~T.~Boothroyd, and C.~R.~M.~Grovenor, Supercond. Sci. Technol. {\bf 25} 084023 (2012). 

\bibitem{Ryu2011} H.~Ryu, H.~Lei, A.~I.~Frenkel, and C.~Petrovic, Phys. Rev. B {\bf 85}, 224515 (2012).

\bibitem{Ozaki2011} T.~Ozaki, H.~Takeya, H.~Okazaki, K.~Deguchi, S.~Demura, Y.~Kawasaki, H.~Hara, T.~Watanabe, T.~Yamaguchi, and Y.~Takano, Europhys. Lett. {\bf 98}, 27002 (2012).

\bibitem{Han2012} F.~Han, H.~Yang, B.~Shen, Z.-Y.~Wang, C.-H.~Li, and H.-H.~Wen, Philos. Mag. {\bf 92}, 2553 (2012).

\bibitem{Lei2011} H.~Lei and C.~Petrovic, Phys. Rev. B {\bf 84}, 212502 (2011).

\bibitem{Krzton2012} A.~Krzton-Maziopa, E.~Pomjakushina, and K.~Conder, to be published in J. Cryst. Growth (2012); doi: 10.1016/j.jcrysgro.2012.01.016.

\bibitem{Wunderlich1990} B.~Wunderlich, Thermal Analysis, New York: Academic Press. pp. 137-140 (1990). 

\bibitem{Suter} A.~Suter and B.~M.~Wojek, Physics Procedia {\bf 30}, 69 (2012).

%

\bibitem{Osborn1945} J.~A.~Osborn, Phys. Rev. {\bf 67}, 351 (1945).

\bibitem{Werthamer} N.~R.~Werthamer, E.~Helfand, and P.~C.~Hohenberg, Phys. Rev. {\bf 147}, 295 (1966).

\bibitem{Bukowski2009} Z.~Bukowski, S.~Weyeneth, R.~Puzniak, P.~Moll, S.~Katrych, N.~D.~Zhigadlo, J.~Karpinski, H.~Keller, and B.~Batlogg, Phys. Rev. B {\bf 79}, 104521 (2009).

\bibitem{Amato} A.~Amato, Rev. Mod. Phys. {\bf 69} 1119 (1997).

\bibitem{Ambegaokar1963} V.~Ambegaokar and A.~Baratoff, Phys. Rev. Lett {\bf 10}, 486 (1963).

\bibitem{Ebner1985} C.~Ebner and D.~Stroud, Phys. Rev. B {\bf 31}, 165 (1985).

\bibitem{Clem1988} J.~R.~Clem, Physica C {\bf 153-155}, 50 (1988). 

\bibitem{Charnukha2012_b} A.~Charnukha, A.~Cvitkovic, T.~Prokscha, D.~Pr\"opper, N.~Ocelic, A.~Suter, Z.~Salman, E.~Morenzoni, J.~Deisenhofer, V.~Tsurkan, A.~Loidl, B.~Keimer, and A.~V.~Boris, Phys. Rev. Lett. {\bf 109}, 017003 (2012).

\bibitem{Torchetti2011} D.~A.~Torchetti, M.~Fu, D.~C.~Christensen, K.~J.~Nelson, T.~Imai, H.~C.~Lei, and C.~Petrovic, Phys. Rev. B {\bf 83}, 104508 (2011).

\bibitem{Charnukha2012} A.~Charnukha, J.~Deisenhofer, D.~Pr\"opper, M.~Schmidt, Z.~Wang, Y.~Goncharov, A.~N.~Yaresko, V.~Tsurkan, B.~Keimer, A.~Loidl, and A.~V.~Boris, Phys. Rev. B {\bf 85}, 100504(R) (2012).

\bibitem{Homes2012} C.~C.~Homes, Z.~J.~Xu, J.~S.~Wen, and G.~D.~Gu, (unpublished) arXiv/condmat:1208.2240v1 (2011).

\bibitem{Simonelli2012} L.~Simonelli, N.~L.~Saini, M.~M.~Sala, Y.~Mizuguchi, Y.~Takano, H.~Takeya, T.~Mizokawa, and G.~Monaco, Phys. Rev. B {\bf 85}, 224510 (2012)

\bibitem{Bianconi1996} A.~Bianconi, N.~L.~Saini, T.~Rossetti, A.~Lanzara, A.~Perali, M.~Missori, H.~Oyanagi, H.~Yamaguchi, Y.~Nishihara, and D.~H.~Ha, Phys. Rev. B {\bf 54} 12018 (1996).

\bibitem{Moon2010} C.~Y.~Moon and H.~J.~Choi, Phys. Rev. Lett. {\bf 104} 057003 (2010).

\bibitem{Das2011} T.~Das and A.~V.~Balatsky, Phys. Rev. B {\bf 84} 014521 (2011).

\bibitem{Andersen2011} O.~K.~Andersen and L.~Boeri, Ann. Phys. {\bf 523} 8 (2011). 

\bibitem{Wittlin2012} A.~Wittlin, P.~Aleshkevych, H.~Przybylinska, D.~J.~Gawryluk, P.~Dluzewski, M.~Berkowski, R.~Puzniak, M.~U.~Gutowska, and A.~Wisniewski, Supercond. Sci. Technol. {\bf 25}, 065019 (2012).

\bibitem{Kresin2012} V.~Kresin, J. Supercond. Nov. Magn. {\bf 25} 711 (2012).









%
%
%
%
%
%
%
%






























\end{thebibliography}
\end{document}